# High Performance Printed AgO-Zn Rechargeable Battery for Flexible Electronics

Lu Yin[1,4], Jonathan Scharf[1,4], Jessica Ma[1], Jean-Marie Doux[1], Christopher Redquest[1], Viet Le[1], Yin Yijie[1], Jeff Ortega[3], Xia Wei[3], Joseph Wang[1,2]*, Ying Shirley Meng[1,2,5]*

[1] Department of Nanoengineering, University of California San Diego, La Jolla, CA 92093, USA
[2] Sustainable Power and Energy Center (SPEC), University of California San Diego, La Jolla, CA 92093, USA
[3] ZPower LLC, Camarillo, CA 93012, USA
[4] These authors contributed equally to this work
[5] Lead Contact
*Correspondence: josephwang@ucsd.edu (J. W.), shmeng@ucsd.edu (Y. S. M.)

**Summary**

The rise of flexible electronics calls for cost-effective and scalable batteries with good mechanical and electrochemical performance. In this work, we developed printable, polymer-based AgO-Zn batteries that feature flexibility, rechargeability, high areal capacity, and low impedance. Using elastomeric substrate and binders, the current collectors, electrodes, and separators can be easily screen-printed layer-by-layer and vacuum-sealed in a stacked configuration. The batteries are customizable in sizes and capacities, with the highest obtained areal capacity of 54 mAh/cm$^2$ for primary applications. Advanced micro-CT and EIS were used to characterize the battery, whose mechanical stability was tested with repeated twisting and bending. The batteries were used to power a flexible E-ink display system that requires a high-current drain and exhibited superior performance than commercial coin-cell batteries. The developed battery presents a practical solution for powering a wide range of electronics and holds major implications for the future development of practical and high-performance flexible batteries.

## Introduction

Recent interest in multifunctional flexible electronics for applications in sensing, displays, and wireless communication advocates for the development of complementary flexible energy storage solutions.[1,2] Accordingly, numerous efforts have been made to tackle the challenges of fabricating batteries with robust mechanical resiliency and high electrochemical performance. [ref] To do so, several studies focused on adapting novel battery fabrication techniques, such as substrate pre-stretching, textile-embedding, and "island-bridge", wire/cable, kirigami, and origami structuring to endow structural flexibility and stretchability to batteries.[3–9] Others studies focused on tackling the challenge by the means of material innovation, developing polymer-based current collectors, electrodes, separators, and electrolytes for various intrinsically flexible and stretchable batteries.[10–17] However, due to the exponential growth in the wearable flexible electronics market, manufacturers started to realize the urgent need for scalable, low-cost, and high-performance flexible battery technologies to provide practical energy storage solutions for the tens of millions of devices produced every year. Many flexible batteries rely on fabrication processes that are complex, low throughput, and high cost, and thus have limited practicality which hinders their lab-to-market transformation. Addressing the need for flexibility and scalability while maintaining low cost, printed high-performance batteries are crucial for realizing most of the commercially viable battery technology. Using low-cost thick-film fabrication technologies, flexible battery components can be printed sheet-to-sheet or roll-to-roll using traditional, low-maintenance screen printing or doctor blade casting equipment, thus realizing low-cost mass production of flexible batteries.[18]

Among many commercialized printed flexible batteries, aqueous zinc (Zn)-based conversion cells were successful in developing products with high throughput and low production cost.[19] The Zn anode chemistry has been of special interest for the flexible battery market due to its low material cost, high theoretical capacity (820 mAh/g, 5854 mAh/L), good rechargeability, and safe chemistry.[20,21] In addition, as Zn and the aqueous electrolyte can be readily handled in an ambient environment, the equipment and production costs of Zn-based batteries are often considerably lower compared to lithium-ion batteries. However, commercial Zn-based printed flexible batteries are usually non-rechargeable and feature low capacity and high impedance, thus limiting their application in low-power, disposable electronics only. Addressing these limitations, several studies have reported the development of printable, rechargeable, and high-performance Zn-based batteries.[16,22,23] Among them, the silver oxide-zinc ($Ag_2O$-Zn) battery has attracted particular attention due to its rechargeable chemistry and its tolerance to high-current discharge.[24,25] The redox reaction relies on the dissolution of zinc ions ($Zn^{2+}$) and silver ions ($Ag^+$) in the alkaline electrolyte and their supersaturation-induced precipitation, which takes place rapidly while maintaining a stable voltage at 1.56 V (**Equation 1-6**). [26,27]

Anode:

(Dissolution) $Zn\ (s) + 4OH^-\ (aq) \leftrightarrow Zn(OH)_4^{2-}\ (aq) + 2e^-$ (1)

(Relaxation) $Zn(OH)_4^{2-}(aq) \leftrightarrow ZnO\ (s) + H_2O\ +\ 2OH^-\ (aq)$ (2)

(Overall) $Zn\ (s) + 2OH^-\ (aq) \leftrightarrow ZnO + H_2O + 2e^-$ $E^o = -1.22\ \text{V vs. SHE}$ (3)

Cathode:

(Dissolution) $Ag + OH^- \leftrightarrow AgOH(aq) + e^-$ (4)

(Relaxation) $2AgOH(aq) + 2OH^- \leftrightarrow Ag_2O(s) +\ H_2O$ (5)

(Overall) $Ag_2O(s) +\ H_2O\ \leftrightarrow 2Ag(s) + 2OH^-(aq)$ $E^o = +0.34\ \text{V vs. SHE}$ (6)

Most of these batteries rely solely on the use of the lower oxidation state of silver to obtain reversible redox reaction, while the higher oxidation state (AgO), with its redox reaction described in **Equation 7**, has been rarely utilized.

$$2AgO\ (s) + H_2O\ + 2e^- \leftrightarrow Ag_2O\ (s) +\ H_2O\ (l) \qquad E^o = +0.60\ \text{V vs. SHE} \quad (7)$$

The underutilization of AgO can be attributed to its instability, namely, its lattice phase change when transitioning into $Ag_2O$, which may result in irreversible shape changes that impede rechargeability, and its high charging potential responsible of possible electrode gassing due to oxygen evolution reaction.[8,28–31] However, once addressing these issues, it is possible to access a much higher theoretical cathode capacity (from 231 mAh/g for $Ag_2O$ to 432 mAh/g for AgO). So far, printed silver-zinc batteries reported in the literature still have low rechargeability (< 50 cycles), limited capacity (< 12 mAh/$cm^2$ for primary cell, < 3 mAh/$cm^2$ for secondary cell), along with high internal resistance (~$10^2$ **Ω**) that results in large voltage drop during operation.[10,29,32–34] Such limitations are hindering the adaptation of silver-zinc printed batteries in flexible electronics.

Herein, we present a novel fabrication process of all-printed, flexible, and rechargeable AgO-Zn batteries with ultra-high areal capacity, low impedance, and good rechargeability as a practical energy storage solution for flexible electronics. The fabrication of the cell relies on low-cost, high-throughput, layer-by-layer printing of formulated powder-elastomer composite inks to form the current collectors, Zn anode, AgO cathode, and their corresponding separators. The battery adopts a low-footprint stacked configuration, with potassium hydroxide (KOH) - poly(vinyl alcohol) (PVA) hydrogel as a low impedance electrolyte sandwiched between the two fully printed electrodes. Using the thermoplastic styrene-ethyl-butylene-styrene block copolymer (SEBS) elastomer-based substrate, the assembled battery can be directly heat- and vacuum-sealed to preserve the electrolyte and ensure appropriate cell pressure during operation. This fabrication and assembly process can be applied to different cell sizes with adjustable areal capacity, allowing customizable battery form factors that are tailored for specific applications. Fully utilizing the higher oxidation state of the AgO, the as-printed cells were able to reach a high areal capacity of > 54 $mAh/cm^2$ while maintaining a low internal resistance (~10 Ω) for primary applications. Furthermore, utilizing an optimized cycling profile, the printed cells were recharged for over 80 cycles, sustaining 0.2 C – 1 C discharges without exhibiting significant capacity loss, while maintaining low impedance throughout each cycle. Moreover, the fabricated cells displayed outstanding robustness against repeated bending and twisting deformations. To demonstrate their performance in powering typical flexible electronics, the fabricated thin-film batteries were successfully implemented in a flexible E-ink display system with an integrated microcontroller unit (MCU) and Bluetooth modules that require pulsed high-current discharge. Leveraging a low-cost scalable production process, polymer-based flexible architecture, and customized ink formulations, the all-printed AgO-Zn battery, with its desirable mechanical and electrochemical performance, presents a practical solution for powering the next-generation flexible electronics, and sets a new benchmark for the further development of printable flexible batteries.

## Results

All-printed Polymer-based Fabrication

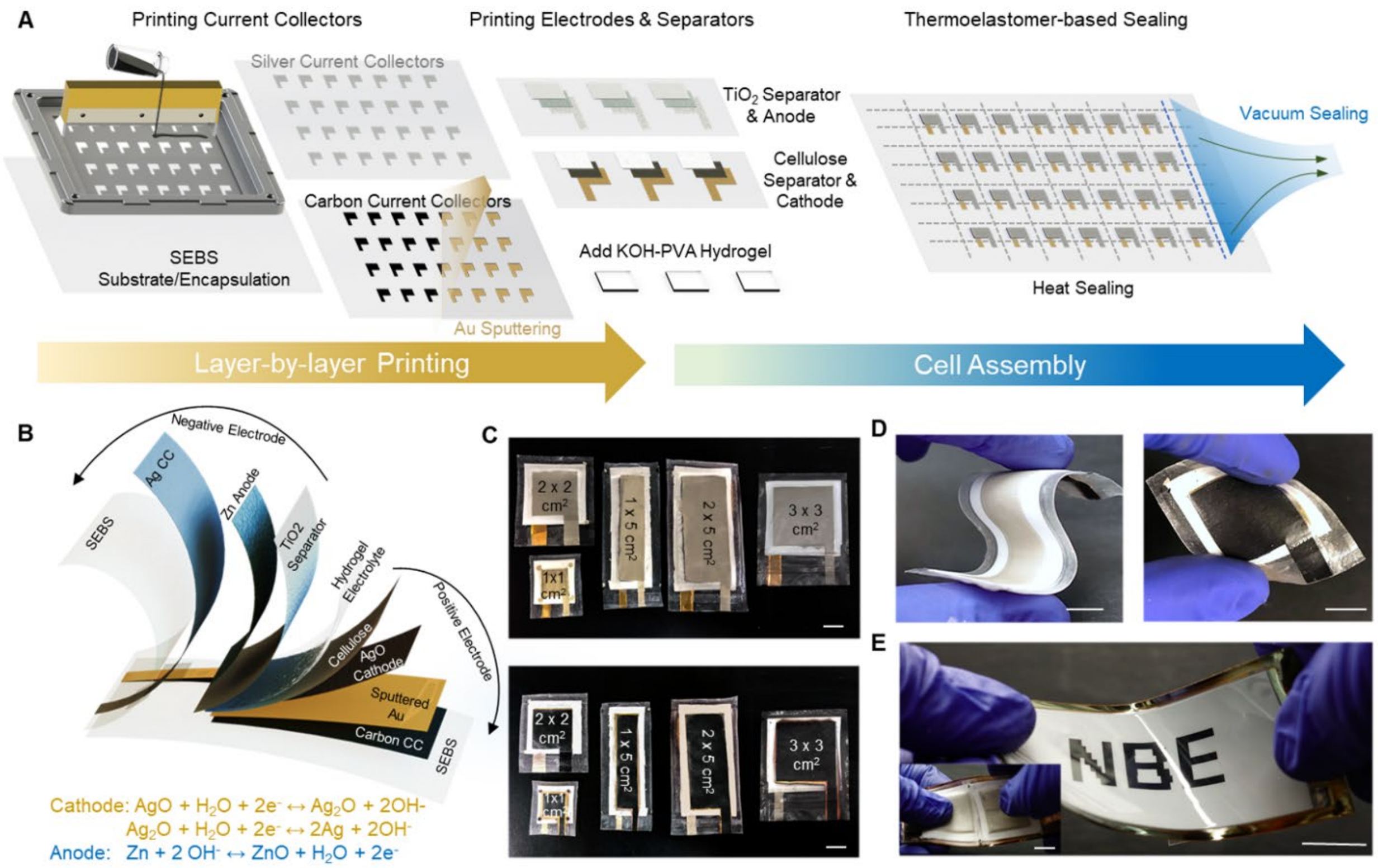

***Figure 1. All-printed fabrication of the flexible, rechargeable, and high-capacity AgO-Zn battery.*** *(A) Illustration of the layer-by-layer printing and vacuum sealing assembly processes. (B) Illustration of the AgO-Zn battery cell structure. The cell is composed of a hydrogel electrolyte sandwiched between the 2 electrodes, with each side composed of a heat-sealable SEBS substrate, current collectors, active material electrodes, and corresponding separators. (C) Photo images of the assembled cells in different customized sizes. (D) Photo images demonstrating the flexibility of the printed batteries. (E) Photo images of a flexible E-ink display system powered by the flexible AgO-Zn batteries. Scale bar, 1 cm.*

The all-printed fabrication of the flexible AgO-Zn battery was designed based on the careful selection of elastomers for the substrate, sealing, and ink binders based on their mechanical properties, chemical stabilities, and processabilities. SEBS was selected as the substrate material for its good solvent processability, chemical stability under high pH, outstanding elasticity, as well as its appropriate melting point (~200 °C), allowing it to be easily cast into films that are chemically stable, flexible, and heat-sealable to support and seal the battery.[35] Screen-printing, a low-cost high-throughput thick-film technique was used for ink deposition, as it allows the efficient fabrication of the current collectors, electrodes, and separators into their preferred shapes and thicknesses.[18] The screen-printing of the batteries relies on the customized formulation of 6 inks corresponding to the current collectors, electrodes, and the separators for both the anode and cathode. Conductive and flexible silver ink and carbon ink were printed as the anode and cathode current collectors respectively. Both inks use SEBS as the elastomer binder and toluene as the solvent to allow the ink to firmly bond to the toluene-soluble SEBS substrate. The anode ink was composed of Zn particles with bismuth oxide ($Bi_2O_3$) as an additive to reduce dendrite formation and suppress $H_2$ gassing, while the cathode ink was mainly composed of AgO powder with a small amount of lead oxide coating to enhance the electrochemical stability and carbon black added to enhance the electronic conductivity of the electrode.[29] A high-pH stable, elastomeric fluorocopolymer was used as the binder for both electrodes for its solubility in lower ketones which is less prone to oxidation by the highly oxidative AgO. Cellulose powder was used to form the porous cathode separator that mimics the use of cellophane to capture and reduce dissolved silver ions and prevent material crossover.[8,19] A titanium dioxide ($TiO_2$)-based ink was formulated for the anode separator, acting as a physical barrier to Zn dendrite growth. Lastly, a solid-phase polyvinyl alcohol (PVA) hydrogel crosslinked with potassium hydroxide (KOH) was prepared as the electrolyte, which complements the cell flexibility without the risk of leaking. Lithium hydroxide (LiOH) and calcium hydroxide ($Ca(OH)_2$) were used as additives in the electrolyte to maintain electrolyte chemical stability and minimize zinc dissolution.[36,37]

The fabrication of the batteries begins with the preparation of the substrates, where a resin of SEBS dissolved in toluene was cast onto wax papers using film casters and dried in the oven to form a transparent elastic film. The layer-by-layer printing process is illustrated in **Figure 1A**. Firstly, the Ag and the carbon inks were printed onto the SEBS substrate as current collectors, with a 400 nm layer of gold sputtered onto the carbon current collectors to enhance their conductivity and chemical stability. Then, the Zn and the $TiO_2$ inks, and the AgO and the cellulose inks were printed onto their corresponding current collectors. To complete the cells, the KOH-PVA hydrogel electrolyte was cut to size and sandwiched between the two electrodes. Lastly, the sheet of batteries was heat and vacuum sealed and separated into individual cells, finalizing the scalable sheet-by-sheet fabrication of multiple cells in one sitting. The flexible, vacuum-sealed AgO-Zn batteries comprised of 9 layers of composite materials, can thus be easily fabricated using layer-by-layer screen-printing (**Figure 1B**). The major advantage of the stencil printing technique is the customizable dimension of the cells that can be tailored for different applications with specific form factor and capacity requirements. As examples, cells in different sizes were fabricated using the same fabrication process (**Figure 1C**), and could be integrated with different sizes of flexible electronic devices. Regardless of the shapes and sizes, the assembled cells are highly flexible and durable under repeated mechanical deformations (**Figure 1D**), making them highly suitable for powering wearable and flexible electronics that require high resiliency to various deformations. Furthermore, the superior electrochemical performance of the fabricated AgO-Zn battery greatly expands the application of thin-film batteries in electronics with high power demands. This capability was demonstrated by powering a flexible display system with microcontroller and Bluetooth modules (**Figure 1E**), as discussed in the later section.

Microstructural and Electrochemical Characterization

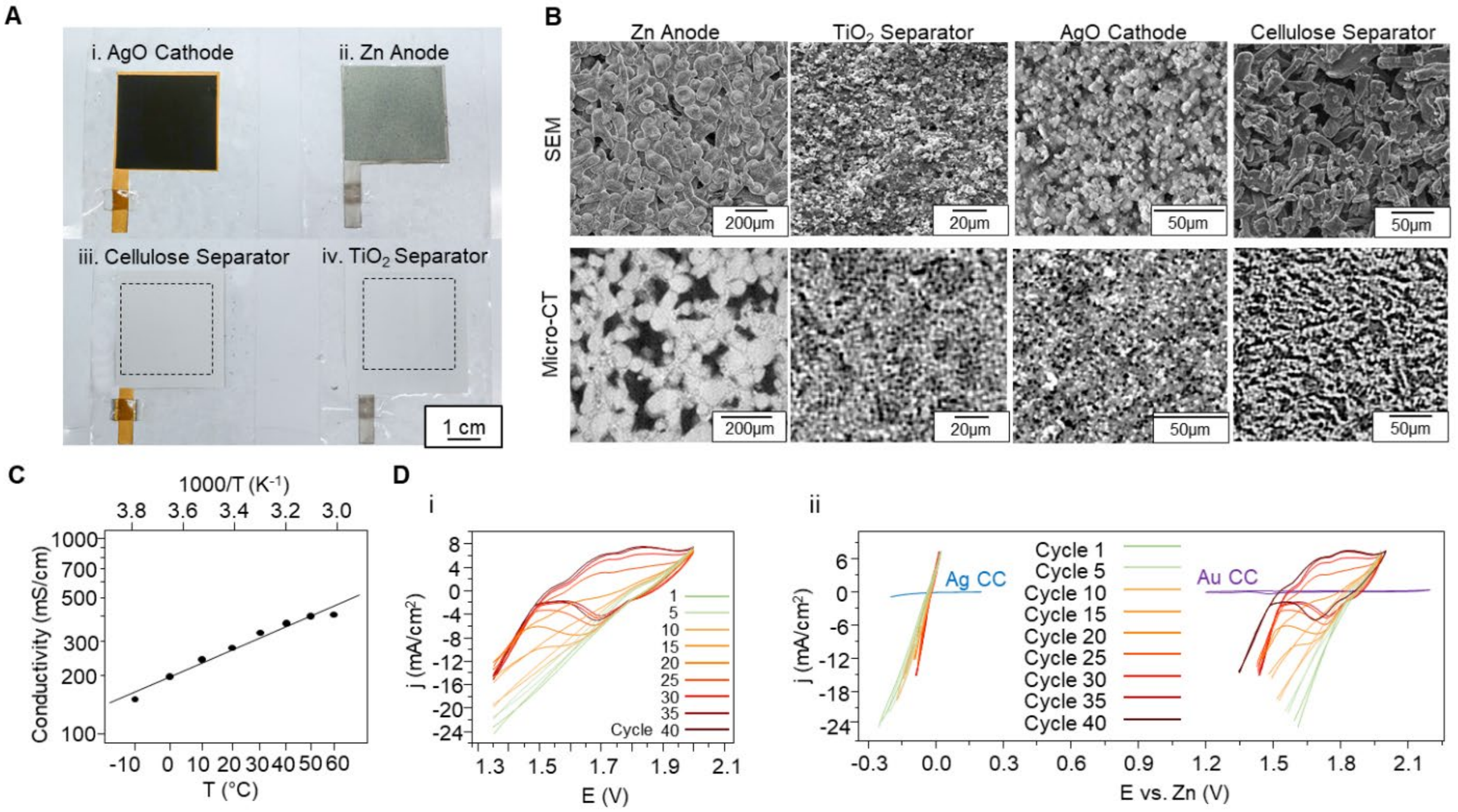

***Figure 2. Morphological and electrochemical characterization of the printed battery.*** *(A) Photo images of the printed 3 × 3 cm² cell with the (i) AgO electrode, (ii) Zn electrode, (iii) Cellulose separator, and (iv) TiO₂ separator layers. (B) Microscopic images of corresponding layers of the cell taken via SEM and Micro-CT. (C) The conductivity of the gel electrolyte as a function of temperature. (D) 40 cycles of CV between 2 V and 1.35 V of the (i) full cell and (ii) corresponding potential shifts in the anode (left) and the cathode (right) using a 3-electrode cell with a Zn metal pseudo-reference electrode. The CVs of the current collectors within the corresponding voltage windows (anode -0.3 V – 0.3 V, cathode 1.2 V – 2.2 V) under the electrolyte environment are overlaid onto the electrode CVs. Scan rate: 10 mV/s.*

The printed electrodes and separators (**Figure 2A**) were characterized by scanning electron microscopy (SEM), as well as non-intrusive, in-situ micrometer-scale X-ray computed tomography (micro-CT). The introduction of micro-CT enables the capability of non-destructive inspection of the battery, which can be highly beneficial to characterize the devices under deformation without the need to disassemble the cells. As shown in **Figure 2B**, the micro-CT images show a morphology which in agreement with the SEM images of the pristine anode, cathode, cellulose separator, and $TiO_2$ separator. Accordingly, the 3-dimensional (3D) imaging of these films shown in **Figure S21** offers gives a more comprehensive understanding of the material structures. The loosely packed Zn anode used in this work is made of large particles, with sizes in the range of 50 - 100 μm, which hence reduces the surface passivation induced by the spontaneous reaction with the electrolyte. Energy Dispersive X-Ray Analysis (EDX) further shows the homogeneous coverage of the $Bi_2O_3$ and the fluoropolymer binders on the surfaces of the Zn particles (**Figure S5**). The $TiO_2$ separator contains much smaller particles to form a dense and homogenous film, thus can effectively reduce the dendrite growth (**Figure S7**). In comparison, the AgO electrode uses 1 - 20 μm particles to produce a porous electrode, which was paired with a separator with similar particle sizes to capture the dissolved Ag species (**Figure S6** and **S8**). Overall, the porous electrodes grant easy permeation of the electrolyte, thus allowing the fabrication of cells with thicker electrodes to increase areal capacity. The conductivity of the PVA-based electrolyte (**Figure 2C**) is in the $10^2$ mS/cm order in a wide range of temperatures (-10 °C to 60 °C), similar to that of other gel electrolytes reported in the literature.[15,38] The solid-phase hydrogel holds the ability to properly wet the electrodes which allows higher current cycling, while serving as a leak-free electrolyte barrier blocking dendrite growth. The hydroxide concentration was shown to have little effect on the electrolyte conductivity (**Figure S9**), but had a significant impact to the cycle life of the battery (**Figure S13**), and was thus optimized to be 36.5 % by weight.

**Figure S10** displays 3-electrode cells, using a Zn foil as a pseudo-reference electrode, that was used for cyclic voltammetry (CV) analysis. The AgO-Zn battery is designed to charge and discharge within the window of 1.35 V to 2 V which is used as the CV scanning range. As shown in the full cell CV in **Figure 2D-i**, within the scanning rate of 10 mV/s, the cell can undergo a high current density of up to 20 $mA/cm^2$, proving the cell's ability to discharge at high current. Using the external Zn reference, the full cell CV can be used to gauge the potential shifts of each electrode separately. As shown in **Figure 2D-ii**, the relative anode potential (left) does not shift significantly during the sweep, whereas the cathode potential (right) contributes to the majority of the potential change in the cell, suggesting that the AgO cathode is being the rate-limiting electrode in the charge-discharge process. The CV of the current collectors in the corresponding voltage window (**Figure S11**) is overlaid in **Figure 2D-ii**, demonstrating the electrochemical stability of the current collectors within the expected potential range. It is worth noting that the current density of the Ag current collector increases towards the negative potential direction, which corresponds to the possible hydrogen evolution reaction taken place on the anode during the charging process. Such undesirable reaction is generally avoided as lower current density is used in the normal charging processes, corresponding to lower anode polarization (**Figure S18**).

<ins>AgO-Zn Battery as A High Areal Capacity Primary Cell</ins>

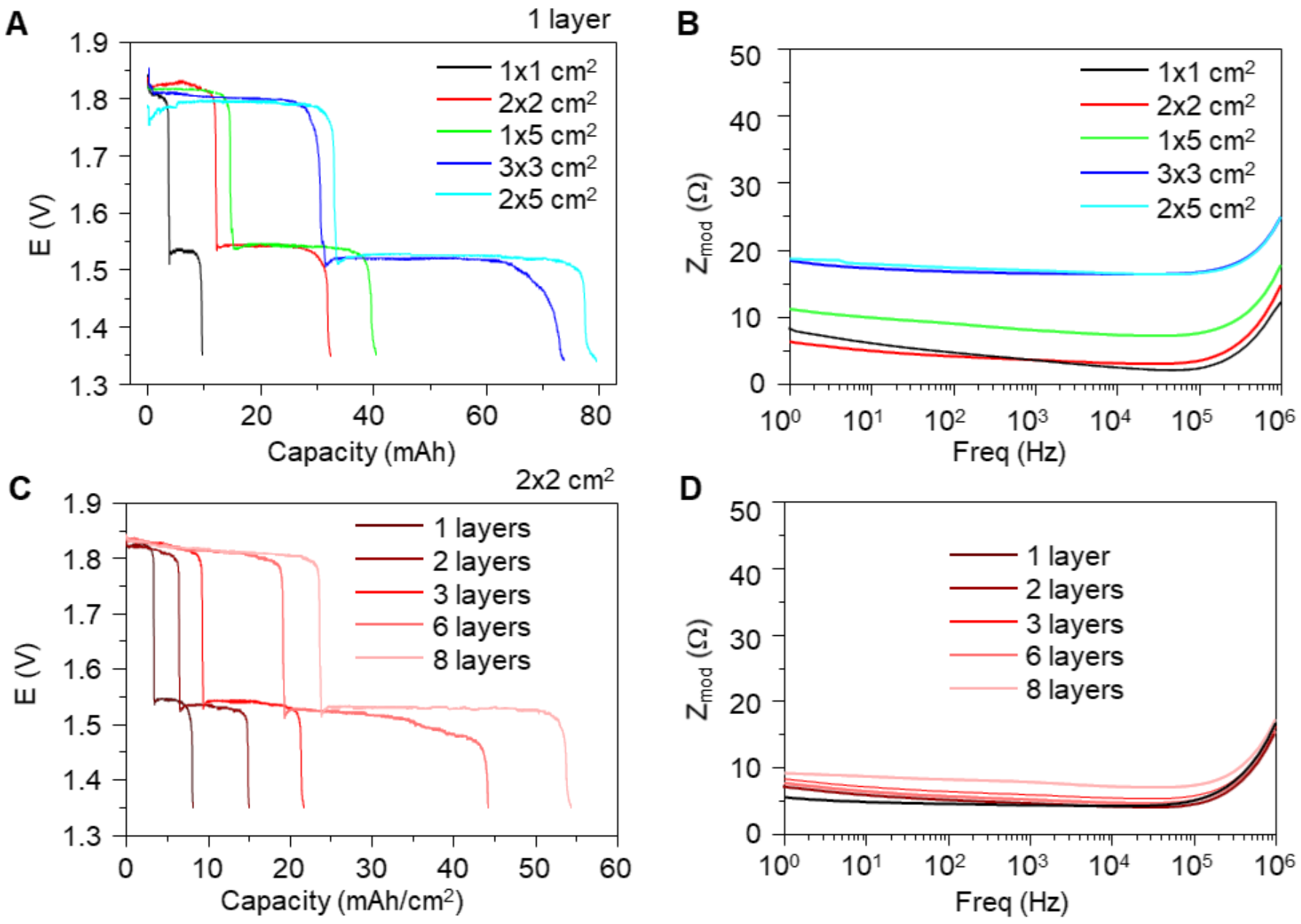

***Figure 3. Electrochemical performance of the AgO-Zn cells as primary batteries.*** *(A) The obtainable capacity of various sizes of cells that were printed with 1 layer of active materials, discharged at a current of 1 mA. (B) Bode plot reflecting the corresponding impedance of cells of different sizes. (C) The obtainable capacity of the 2 × 2 cm² cells with active material loading from 1 layer to 8 layers. (D) Bode plot reflecting the corresponding impedance of the 2 × 2 cm² cells with different areal loading.*

The ability of the cell design to adapt to different cell sizes and areal loadings was evaluated. Cells with the same electrode thickness but different form factors, by varying the electrode designs, as well as the cells with the same form factors and different thicknesses by varying the number of layers of active material printed, were fabricated and discharged at a constant 1 mA current. As shown in **Figure 3A**, cells with 1-layer (anode ~ 120 µm, ~ 45 mg/cm$^2$, cathode ~ 75 µm, ~ 26 mg/cm$^2$) of electrode thickness with the sizes of 1 × 1 cm$^2$, 2 × 2 cm$^2$, 1 × 5 cm$^2$, 2 × 5 cm$^2$ and 3 × 3 cm$^2$ were prepared, and the capacity increases proportionally to the cell area, with an average areal capacity of 8 mAh/cm$^2$. The impedance of these cells was measured via 2 electrodes EIS, presented in **Figure 3B**. The overall increase in impedance throughout the high frequency and low-frequency domain suggests an increase in cell contact resistance, caused by the increase in resistance of the current collector as the cell size increases. Cells with a size of 2 × 2 cm$^2$ were also characterized with increasing areal loadings by printing 1, 2, 3, 6, and 8 layers of electrodes. As demonstrated in **Figure 3C**, as the areal loading of active material increases, the areal capacity of the cell increases proportionally, reaching as high as 54 mAh/cm$^2$ with 8 layers of electrodes (anode ~ 800 µm, ~ 310 mg/cm$^2$, cathode ~500 µm, ~ 180 mg/cm$^2$). The EIS on the cells with different thicknesses also showed no significant impedance increase as the thickness increases: only a minor increase in impedance in the low-frequency domain suggests a slight increase in the diffusion resistance due to thicker electrodes (**Figure 3D**). Such behavior can be attributed to the large pore sizes in both the anode and the cathode, which cause little resistance for the ion diffusion. Overall, the printed AgO-Zn cell was able to uphold superior performance in a wide range of sizes and areal loading, thus proven its customizability as a primary thin-film battery to power various electronics with appropriate sizes and capacity. A comparison between this and other flexible batteries is shown in **Table S1** and **Figure S1**, showing the obtained areal capacity of 54 mAh/cm$^2$ being the highest among all printed batteries.

AgO-Zn Battery as High-performance Secondary Cell

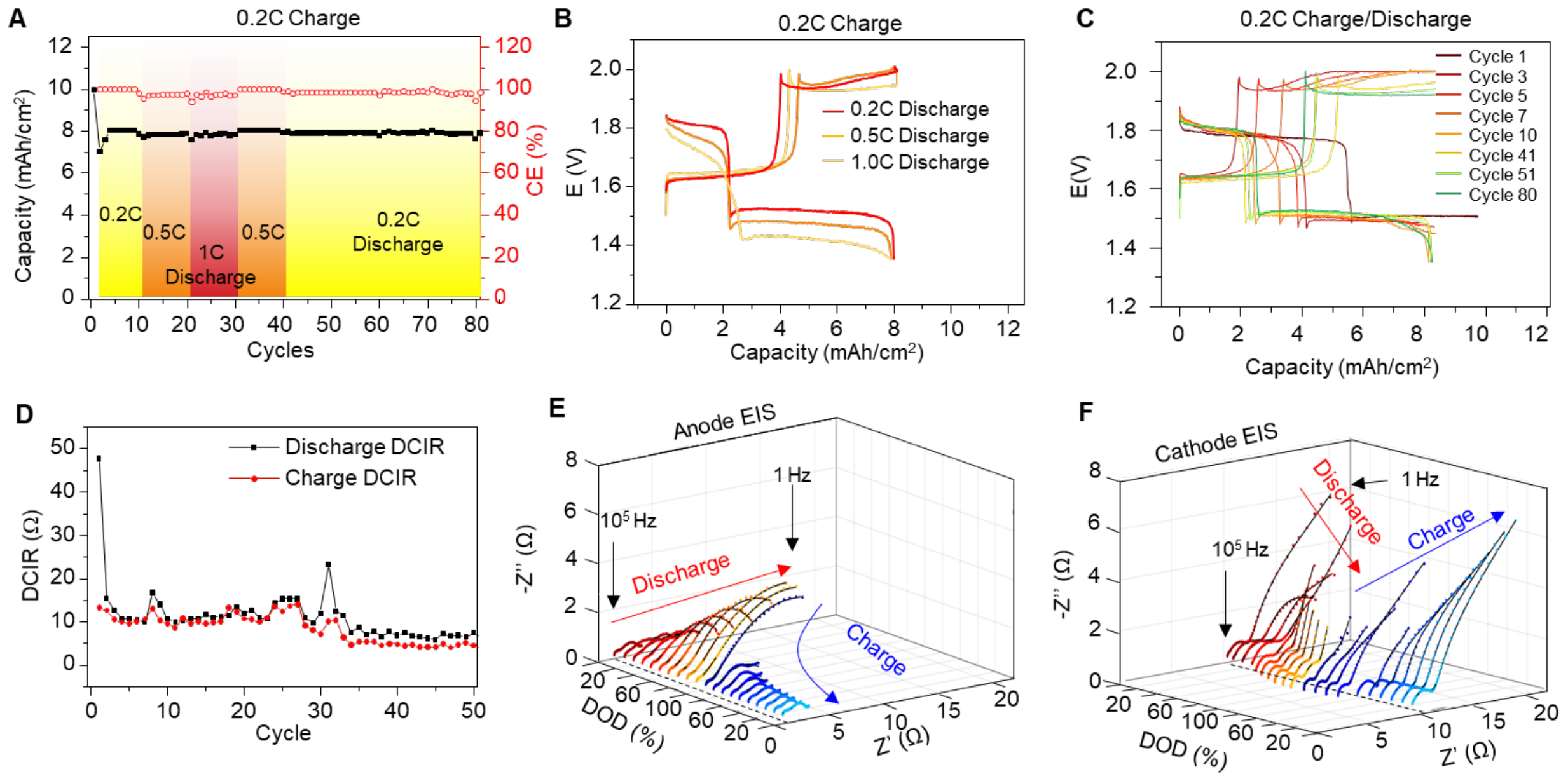

***Figure 4. Electrochemical performance of the AgO-Zn cells operating as rechargeable batteries.*** *(A) The cycling performance of the printed battery with a charging C-rate of 0.2C and varying discharge rate of 0.2C, 0.5C, and 1C. (B) The voltage-capacity plot of the battery under different discharging C-rates. (C) The voltage-capacity plot of the AgO-Zn battery at different number of cycles showing the stabilization of the charge-discharge profile. (D) The DCIR of the AgO-Zn within 50 cycles cycled at the C-rate of 0.2C. The EIS profile of the (E) Zn anode and (F) AgO cathode within 1 complete discharge-charge cycle on a 3-electrode cell with a Zn metal pseudo-reference electrode.*

Beyond the application as a primary battery, the electrochemical performance of the flexible AgO-Zn battery as a secondary cell was also characterized. As a cell operating with conversion-type chemistry, it is crucial to avoid over-oxidation of the anode materials or over-reduction of the cathode material that would lead to irreversible particle shape change. Previous studies report that the loss of capacity in this system is due to the increased thickness of the ZnO layer that passivates the anode surface, as well as the coarsening of the AgO/$Ag_2O$ particles leading to a decrease in cathode surface area.[8,36] Such behavior can be effectively mitigated by accurately controlling the degree of charge and discharge to limit the occurrence of irreversible electrode shape changes. The optimized charge-discharge algorithm was determined to cycle the cell between 40% and 90% of its maximum capacity, with larger ranges resulting in lower cycle life as shown in **Figure S12**. **Figure 4A** demonstrates the cycling of a cell with 2-layer electrodes with a maximum capacity of ~ 16 mAh/cm$^2$. A formation cycle is firstly performed, discharging 10 mAh/cm$^2$ (60% of max. areal capacity) at the rate of 0.1C, allowing the electrode to slowly relax into its preferred morphology with increased surface area and reduced impedance. Then, the batteries were charged at 0.2 C rate until reaching 2 V and charged at constant voltage until the C-rate dropped to below 0.04 C or the capacity reached 8 mAh/cm$^2$ (50% of max. areal capacity). The battery was then discharged at 0.2 C until reaching a columbic efficiency of 100% or a voltage of 1.35V. The entire charge-discharge process is accurately controlled by capacity in the initial cycles, ensuring the cell is cycled between 40% to 90% of its maximum capacity. As shown in **Figure 4C**, after a few cycles at the rate of 0.2 C, the cell slowly relaxed from capacity-controlled discharge to voltage-controlled discharge, with the higher plateau to lower plateau ratio resembling the behavior of the primary cells. Using such charge-discharge algorithm, the cycle life of the unstable AgO oxidation state could be controlled, and a significantly increased cycle life can be obtained compared to previous studies.[10,29,39] Due to the supersaturation-precipitation reaction mechanism of both the anode and the cathode during discharge, the cell can be discharged at a high C-rate of up to 1 C without any loss in capacity and columbic efficiency, as shown in **Figure 4A-B**. Recharging at a higher C-rate is also possible, as shown in **Figure S14**, although this would require a higher capping voltage, reducing the rechargeability and increasing and the risk of oxygen evolution on the cathode, thus was not preferred.

Impedance measurements of the flexible batteries showed relatively low impedances throughout cycling. The impedances of the batteries were either determined during cycling of the full-cell using direct current internal resistance (DCIR) method, or during cycling of the separated anode and cathode half-cells using a 3-electrode configuration with a Zn foil serving as the reference. The DCIR analysis offers a straightforward and simple way to gauge the change in the internal resistance of the battery. As shown in **Figure 4D**, 2-electrode DCIR analysis with both charging and discharging current was performed before each charge and discharge for a battery cycling at 0.2 C, and the battery was able to maintain low internal resistance throughout the cycles, suggesting no formation of high-impedance passivating layers on the surface of the electrode throughout cycling. To obtain detailed information on the change in the impedance of each electrode during each cycle, multiple 3-electrode EIS analyses were performed on the battery while cycling at 0.2 C, and is plotted against the degree of discharge (DOD) of the battery. As presented in **Figure 4E**, the anode half-cell started at a low impedance of 1 – 4 Ω, with 2 depressed semicircles attributed to the high-speed charge transfer at the Zn particle interface and the lower speed hydroxide ions (OH-) diffusion in the porous network.[38] With discharging the low-frequency semicircle slowly expands due to the formation and growth of the ZnO species that impedes the $OH^-$ transport and increases the double-layer capacitance. During charging, the oxygen species are liberated from the reactions in **Equations 1-3** to form $OH^-$ ions that diffuse readily out of the anode. This results in the fast mass transport of $OH^-$ ions out of the anode and a rapid drop in the impedance at the onset of charging that eventually recovers to the initial level, thus showing the reversibility of stripping and depositing of Zn on the anode. For the cathode half-cell EIS showed in **Figure 4F**, at the start of the discharge (0 % DOD), a single semi-circle corresponding to the mass transfer resistance and capacitance of the $Ag_2O$ formation is observed with a low-frequency impedance tail at an angle of approximately 45° suggesting standard Warburg diffusion of $OH^-$. As the cell is discharged, the overall impedance decreases with a second semicircle emerging near the low-frequency domain that can be attributed to the charge transfer resistance and capacitance of Ag formation from $Ag_2O$. During charging, this second low-frequency semicircle disappears as all the Ag oxidizes to form $Ag_2O$ and eventually AgO.

Overall, the 3-electrode impedance results provide a deeper insight into the reaction and possible routes in improving the battery's cycle-life and performance. These data indicate that the impedance of the AgO cathode is responsible for the majority of the cell impedance. Therefore, a beneficial next step will be to incorporate additives that can increase the cathode electrical conductivity to improve the performance in high-current applications. For the anode, the monitoring of ZnO formation via EIS can be paired with topological characterization methods to better control the conversion of Zn electrodes towards extended cycle life and is expected to be extremely useful for future analysis. More data and discussion of 3-electrode EIS can be found in the supplementary section and **Figures S15-17**.

Mechanical Stability of The Flexible, Thin-film Battery

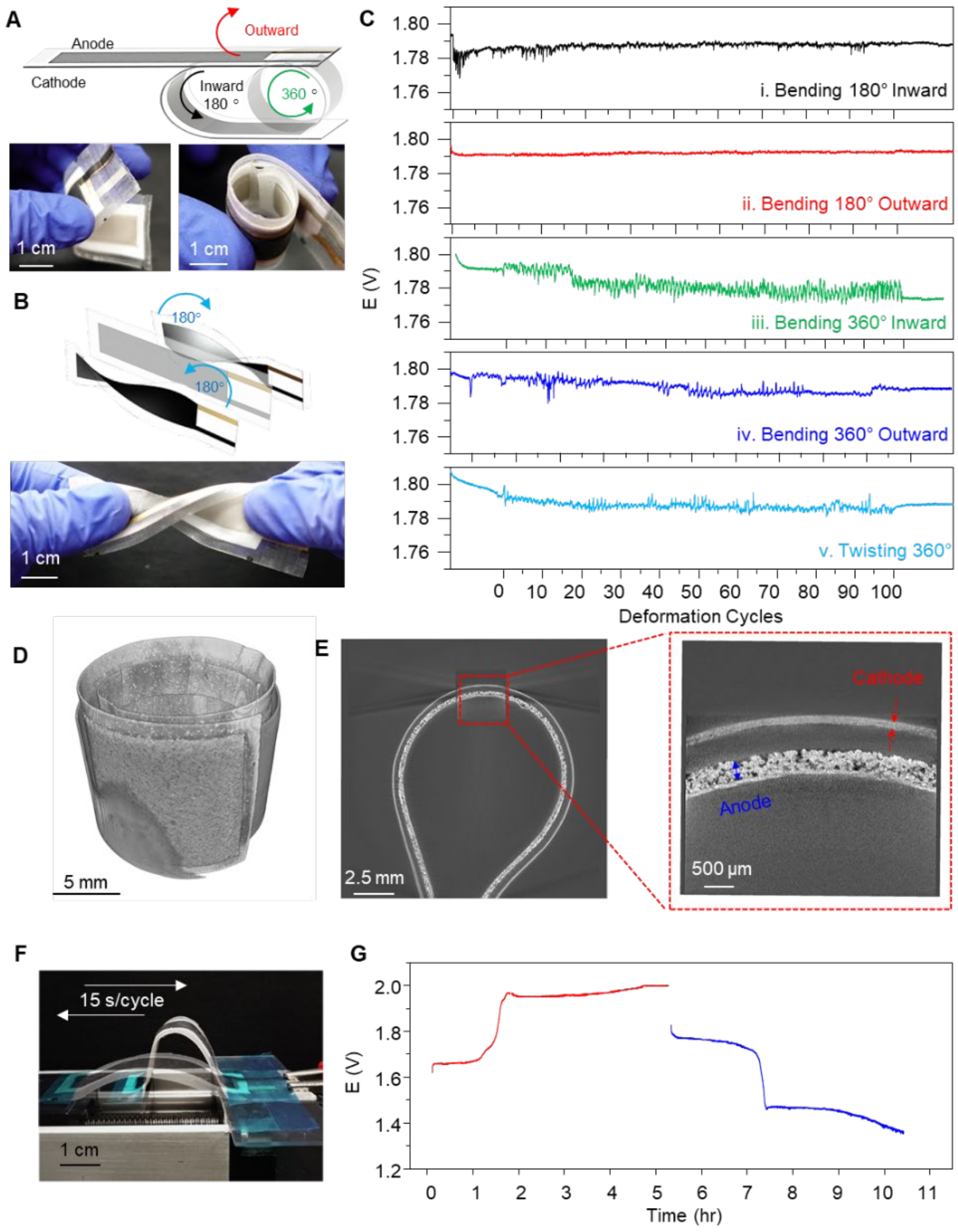

***Figure 5. Performance of the AgO-Zn cell under various mechanical deformations.*** *(A) Illustrations and corresponding photo images of a 2-layer loading, 1 × 5 cm²battery undergoing (A) 180° and 360° bending deformations and (B) 360° twisting deformation. (C) The corresponding voltage profile of the battery during 1 mA discharge while undergoing 100 cycles of (i) 180° outward bending, (ii) 180° inward bending, (iii) 360° inward bending, (iv) 360° outward bending with a bending diameter of 1 cm, and (v) 360° head-to-end twisting. (D) The micro-CT image of the entire 1 × 5 cm² cell after repeated bending and twisting cycles rolled in a diameter of 1 cm, and (E) the cross-section of it bent in a diameter of 1cm (left) and a zoomed-in view (right) of the electrodes, demonstrating no structural damage or delamination of the cell after repeated mechanical deformations. (F) Photo illustration of a battery under repeated 180° bending cycles controlled by a linear stage at the speed of 15 s/cycle, and (G) the corresponding voltage-time plot of the charging (red) and discharging (blue) of the battery during ~2500 repetitions of bending.*

Compared to coin-cell, cylindrical or prismatic cells, the printed flexible batteries have the unique advantage of allowing bending, flexing, and twisting without causing structural failure. To endow such mechanical resiliency, the printed AgO-Zn batteries are composed of flexible and stretchable polymer-particle composite layers endowed the highly elastic binders. This flexibility and stretchability allow the layers to deform to release the inter-layer strain, thus allowing the battery to endure large deformation without delamination between layers or build-up of fatigue, even when very thick electrodes are used. [ref] To test the performance of the batteries under severe strain, a 2-layer 1 × 5 $cm^2$ cell was fabricated and discharged at a current of 1 mA while undergoing repeated bending and twisting deformations. As illustrated in **Figure 5A-B**, the cell was tested with 180° and 360° bending in both directions with a bending radius of 0.5 cm, as well as 180° twisted in both directions from head to end. The corresponding voltage change during 100 cycles (1 s per cycle) of deformation was recorded, as shown in **Figure 5C**. In general, the cell exhibited stable performance during bending and twisting in both directions, with negligible fluctuation in voltage during the 180° bending cycles, and roughly 10 mV fluctuation during the 180° bending and twisting cycles. The inward bending in general shows slightly more variations, which is suspected to be caused by the softer Ag current collector on the anode size undergoing more stretching on the outside during bending. More data of the cell under 10 % stretching deformation can be found in **Figure S20**. To ensure the mechanical stability of the cell, micro-CT was used to characterize the cell after the repeated deformation. As shown in **Figure 5D-E**, the entire cell can be scanned at a high resolution to obtain a 3-dimensional (3D) image reflecting the microscopic structure of the cell under deformation. The zoom-in view of the cross-section of the battery further shows no cracks or delamination after the repeated deformation cycles, reflecting the robust mechanical resiliency of the battery. More 3D visualizations of the battery under bending deformation can be found in **Figure S21**. The rechargeability of the cell is also not interrupted by the repeated deformation, as shown in **Figure 5F-G** and **Video S1**, where the battery can be normally charged and discharged while undergoing ~2500 cycles of 180° bending cycles. Overall, pairing the superior electrochemical and mechanical performance, the printed thin-film AgO-Zn thin-film battery is proven to be well-suited to reliably and sustainably power various wearable and flexible electronics.

Powering Flexible Electronics

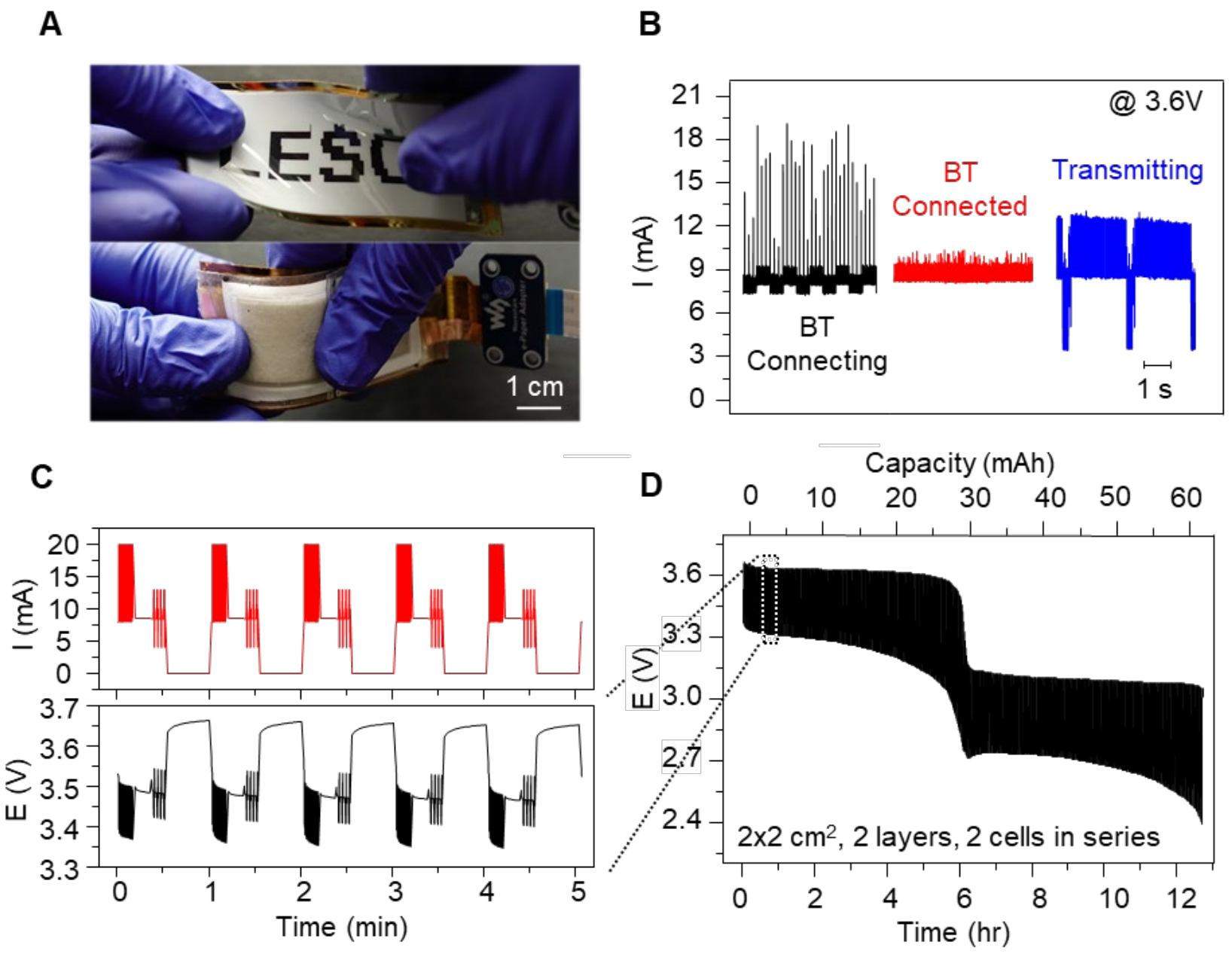

***Figure 6. The powering of a flexible E-ink display system by the flexible AgO-Zn batteries.*** *(A) Photo images of the flexible E-ink display and the placement of two 2-layer loading, 2 × 2 cm² batteries connected in series on the back of the display. (B) The power consumption of the E-ink display system with integrated BT and MCU modules during BT connection (black), after establishing the connection (red), and during active data transmission (blue). (C) Simulated discharge current profile with varying pulses and baselines (red) and the corresponding voltage response of the battery (black). (D) The complete discharge profile of the two cells connected in series implementing the simulation discharge profile.*

To demonstrate the performance of the battery powering typical flexible electronics, we designed a flexible E-ink display system controlled by an Arduino-type microcontroller unit with added Bluetooth (BT) communication module, which resembles many prototypes of IoT, wearable and flexible devices (**Figure 6A and S22**). The system is powered by two 2 × 2 cm$^2$ batteries with 2-layer electrodes connected in series, which can supply enough voltage (>3 V) to boot the system. A mobile device can connect and transmit data and commands to the BT module, which is processed by the microcontroller that refreshes the E-ink display. First, the energy consumption of the system under different operation modes operating at 3.6 V was measured. **Figure 6B** displays the current draw when (1) the system is broadcasting to seek for connection, which contains short bursts of current peaks around 20 mA (in black); (2) the system is connected to a mobile device, with an average current of 9 mA (in red); and (3) the system is actively transmitting data between the cellphone and the display, with the current alternating between a higher baseline of 8.5 mA with peaks of 13 mA, and a lower baseline of 4 mA with peaks of 10 mA (in blue). The batteries are thus discharged using a script simulating the power consumption of the flexible E-ink display system working in repeated discrete sessions, with 10 s of BT broadcasting, 10 s of idle after establishing the connection, 10 s of active data transmission, followed by 30 s of resting (powered off) (**Figure 6C** and **S24**). As illustrated in **Figure 6D**, The two batteries in series were able to sustain the pulsed, high-current discharge in the 3.6 V – 2.4 V window to deliver power to the system constantly for over 12 hours, and is able to maintain its capacity of ~60 mAh, similar to the capacity obtained from the constant low-current 1 mA discharge. By pairing with the high-areal capacity flexible AgO-Zn battery, the flexible E-ink display was able to operate while undergoing bending deformations, which is demonstrated in **Video S2**. In comparison, commercial lithium coin cells with similar rated capacity were not able to sustain the high current pulsed discharge, resulting in a significant loss in capacity when discharged using the same script (**Figure S24**). The low-impedance and high-energy-density battery is therefore proven to have both outstanding electrochemical and mechanical performance for powering of a typical prototype of a flexible electronic system. With their performance even surpassing its non-flexible commercial coin cell counterpart, such all-printed battery can be considered extremely attractive due to its customizability, and flexibility towards real-life applications. A typical application of using the battery to illuminate an LED bulb while applying various mechanical deformations were also tested, where the light intensity does not change as the battery is bent, folded, twisted, and stretched (**Videos S3**).

## Discussion

In this work, we have demonstrated a flexible and high-performance thin-film AgO-Zn battery based on the rechargeable conversion chemistry. Using specially formulated ink with stretchable elastomeric binders and thermoplastic elastomeric substrates, the batteries can be printed layer-by-layer using low-cost, high-throughput screen-printing techniques and assembled with a heat and vacuum sealing processes. To obtain a low device footprint while maintaining easy processability, printable and flexible separators and solid-phase KOH-PVA hydrogel were developed to allow a stacked sandwich configuration. The printable battery is compatible with various cell sizes and areal loading, leading to a high areal capacity of 54 $mAh/cm^2$ in connection to repeated multilayer printing for primary applications. The battery is also rechargeable upon implementing the capacity-controlled cycling algorithm, with high cycle life beyond 70 cycles with varying discharge C-rates without loss in capacity and coulombic efficiency. The battery exhibited low impedance within each discharge-charge cycle, while maintaining low internal resistance throughout multiple cycles, suggesting stable and reversible electrode morphological change during electrode redox reactions. As a flexible energy storage unit for powering various flexible, wearable electronics, the performance of the battery was evaluated under rigorous mechanical testing, demonstrating that the battery offers remarkable resiliency against repeated large deformation bending and twisting cycles. The fabricated batteries were used in the powering of a customized flexible E-ink display system with BT connectivity and delivered an outstanding performance that surpassed commercial coin cells under the high-current pulsed discharge regime required by the electronics. Future work leveraging advanced electrochemical and topological characterization on this promising battery chemistry will be conducted to further improve its performance and cycle life. This will involve further optimization of the fabrication process, the ink composition, and the layer thicknesses and porosities should lead to a practical, commercially viable product with higher cycle life, lower impedance, smaller device footprint, and lower production cost. Overall, this work demonstrates the scalable fabrication of flexible thin-film AgO-Zn batteries with highly desirable electrochemical and mechanical performance and tremendous implications towards the development of novel energy storage devices for the powering of next-generation electronics.

**Experimental Procedures**

Resource Availability

*Lead Contact*

Further information and requests for resources and materials should be directed to and will be fulfilled by the Lead Contact, Shirley S. Meng (shmeng@ucsd.edu).

*Material Availability*

This study did not generate new unique materials.

*Data and Code Availability*

The data of this study are available from the authors upon reasonable request.

Chemicals

$Bi_2O_3$, $Ca(OH)_2$, KOH (pellets, ⩾85%), LiOH, methyl isobutyl ketone (MIBK), toluene, cellulose (microcrystalline powder, 20 μm), Triton-X 114, Poly(ethylene oxide) (PEO) (MW 600,000), and PVA (MW = 89000 – 98000, 99+% Hydrolyzed) were purchased from Sigma Aldrich (St. Louis, MO, USA). Zn, AgO, and $TiO_2$ were obtained from Zpower LLC (Camarillo, CA, USA). The fluorocopolymer (GBR-6005, poly(vinylfluoride-co-2,3,3,3-tetrafluoropropylene)) was obtained from Daikin US Corporation (New York, NY, USA). SEBS (G1645) was obtained from Kraton (Houston, TX, USA). Graphite powder was purchased from Acros Organics (USA). Super-P carbon black was purchased from MTI Corporation (Richmond, CA, USA). All reagents were used without further purification.

Cell Fabrication

*Formulation of The Flexible Inks*

The electrode resin was prepared by adding 5 g of the fluorine rubber in 10 g of MIBK solvent and left on a shake table until the mixture was homogeneous. The SEBS resin was prepared by adding 40 g of the SEBS into 100 mL of toluene and left on a shake table until the mixture was homogeneous.

The silver current collector ink was formulated by combining Ag flakes, SEBS resin, and toluene in 4: 2: 1 weight ratio and mixing in a planetary mixer (Flaktak Speedmixer™ DAC 150.1 FV) at 1800 rotations per minute (RPM) for 5 min.[40] The carbon current collector ink was formulated by firstly mixing graphite, Super-P, and PTFE powder in 84: 14: 2 weight ratio with a set of pestle and mortar. The mixed powder was mixed with the SEBS resin and toluene in a 10: 12: 3 weight ratio using the mixer at 2250 RPM for 10 min to obtain a printable ink.[10]

The Zn anode ink was formulated by firstly mixing the Zn and $Bi_2O_3$ powders in a 9:1 ratio with a set of pestle and mortar until the Zn particles are evenly coated with the $Bi_2O_3$ powder. The evenly mixed powder was then mixed with the electrode resin and MIBK in a 20: 4: 1 weight ratio using the mixer at 1800 RPM for 5 min to obtain a printable ink. The AgO cathode ink was formulated by firstly mixing the AgO and Super-P powders in a 95: 5 weight ratio using a set of pestle and mortar until homogeneous. The powder was then mixed with the electrode resin and MIBK in 5: 5: 2 weight ratio using the mixer at 2250 RPM for 5 min to obtain a printable ink.

The $TiO_2$ separator ink was prepared by firstly mixing $TiO_2$ and cellulose powder in a 2: 1 ratio using a set of pestle and mortar. The mixed powder was then added with the SEBS resin, toluene and Triton-X in 50: 55: 75: 3 weight ratio and mixed with the mixer at 2250 RPM for 10 min to obtain a printable ink. The cellulose separator ink was prepared by firstly mixing $TiO_2$ and cellulose powder in a 26: 9 ratio using a set of pestle and mortar. The mixed powder was then added with the electrode resin, MIBK in an 8: 7: 4 weight ratio and mixed with the mixer at 2250 RPM for 10 min to obtain a printable ink.

*Preparation of The SEBS Substrate*

A resin with 40.8 wt% of SEBS dissolved in toluene is prepared and was left on a linear shaker (Scilogex, SK-L180-E) overnight or until the mixture became transparent and homogeneous. Wax paper was used as the temporary casting substrate, and a film caster with the clearance of 1000 um was used to cast the SEBS resin onto the wax paper. The cast resin was firstly dried in the ambient environment for 1 h, followed by curing in a conventional oven at 80 °C for 1 h to remove the excess solvent. The transparent, uniform SEBS film, which can be readily peeled off from the wax paper after curing, was used as the substrate for subsequent battery printing.

*Printing of The Electrodes*

Stencils for printing the current collectors, electrodes, and separators were designed using AutoCAD software (Autodesk, San Rafael, CA, USA) and produced by Metal Etch Services (San Marcos, CA), with dimensions of 12 in × 12 in. The thickness of the stencils was designed to be 100 µm for the carbon and silver current collectors, 300 µm for the $TiO_2$ separator and the Zn anode, and 500 µm for the cellulose separator and the AgO cathode. To print the anode, the silver ink was first printed onto the SEBS substrate and cured in a conventional oven at 80 °C for 10 min. The Zn ink was then printed onto the silver current collectors and cured at 80 °C for 30 min. The $TiO_2$ ink was lastly printed onto the anode and cured at 80 °C for 10 min. To print the cathode, the carbon ink was firstly printed onto the SEBS substrate and cured at 80 °C for 10 min. PET sheets were cut using a computer-controlled cutting machine (Cricut Maker®, Cricut, Inc., South Jordan, UT, USA) into a mask exposing the printed carbon electrodes, and the masked carbon current collector was sputtered with ~ 400 nm of Au and adhesion interlayer of Cr at a DC power of 100W and 200W respectively and an Ar gas flow rate of 16 SCCM using a Denton Discovery 635 Sputter System (Denton Discovery 635 Sputter System, Denton Vacuum, LLC, Moorestown, NJ, USA). The AgO ink was then printed onto the sputtered current collectors and cured at 50 °C for 60 min. Lastly, the cellulose ink was printed onto the cathode and cured at 50 °C for 60 min. To print multiple layers of electrodes or the separators, the stencil was offset by an additional 65 µm for each layer of AgO and 100 µm for each layer of Zn to compensate for the electrode thickness. See **Figure S2** for the step-by-step printing and assembly process of the batteries and **Figure S3** for the thickness calibration for the printing of anode and cathode.

*Synthesis of The Electrolyte Hydrogel*

The hydrogel is synthesized by mixing the PVA solution and the hydroxide solution into a gel precursor and dried in a desiccator until the desired weight is reached. For synthesizing the 36.5 % hydroxide gel used in this study, the following formulations were used. A hydroxide solution was prepared by dissolving 9.426 g KOH and 0.342 g LiOH into 50 mL deionized (DI) water. 0.5g $Ca(OH)_2$ was then added into the solution and stirred in a closed container under room temperature for 1 h to saturate the solution with $Ca(OH)_2$, and the excess $Ca(OH)_2$ was then removed from the solution. A PVA solution was prepared by dissolving 4.033 g PVA and 0.056 g PEO into 50 mL DI water heated to 90 °C. The precursor solution was prepared by mixing the hydroxide solution and the PVA solution in the weight ratio of 13.677: 10 and poured into a flat petri dish with the weight of 0.2 g/cm$^2$. The precursor was left to dry in a vacuum desiccator until the weight decreased to 26.12 % of precursor weight to obtain a soft, translucent hydrogel with its caustic material taking 36.5 % of the sum of caustic material and the water content. Additional weight and conductivity information for different hydroxide concentrations can be found in **Table S2**. The hydrogel can ben then cut into desired sizes and directly used or stored in a hydroxide solution with the same weight ratio of hydroxide without PVA. The storage solution for the 36.5 % KOH-PVA gel was prepared similar to the hydroxide solution, where 10.777 g KOH, 0.391 g LiOH, and 0.5g $Ca(OH)_2$ were dissolved into 15 mL DI water and the excess $Ca(OH)_2$ was removed. More images of the fabrication of the hydrogel are shown in **Figure S4**.

Microstructural Characterization

Morphological analyses of the current collectors, separators, and active material electrodes were performed with SEM and micro-CT. SEM images were taken using an FEI Quanta FEG 250 instrument with an electron beam energy of 15 keV, a spot size of 3, and a dwell time of 10 µs. Micro-CT experiments were conducted using a ZEISS Xradia 510 Versa. For individual film analysis, micro-CT samples were prepared by punching 2 mm radii disks and stacking them in a PTFE cylindrical tube with alternating PTFE films to provide separation between neighboring film disks. For the Micro-CT of full and sealed cell bending, a 1 × 5 $cm^2$ Zn-AgO battery was bent or rolled around a polyethylene (PE) cylindrical tube with a diameter of 1cm.

For the micro-CT active material electrodes, the heavier metals, such as Zn and Ag, warranted higher X-Ray energies than the printed polymer separator films. Accordingly, scans at 140 keV and a current of 71.26 µA were performed with high energy filters and a magnification of 4X on the Zn and AgO films with voxel sizes of 2.5 µm and 0.75 µm and exposure times of 2 s and 18 s respectively. For the polymer separators, 80 keV scans with an 87.63 µA current were used with low energy filters at a magnification of 4X with voxel resolutions of 0.75 µm and 1.1 µm and exposure times of 8 s and 1 s for the printed anode and cathode separators respectively. For scans of the full cell bending, a voltage of 140 keV and a current of 71.26 µA with a 4X magnification was performed with the following voxel resolutions and exposure times for the respective cases: 18.35 µm and 2 s for low resolution bending scan, 3.54 µm and 5 s for higher resolution bending scan, and 7.55 µm and 2 s for rolled cell scan. For all micro-CT scans conducted, 1801 projections were taken for a full 360˚ rotation with beam hardening and center shift constants implemented during the data reconstruction. Post measurement imaging and analysis were performed by Amira-Avizo using the Despeckle, Deblur, Median Filter, Non-local Means Filter, Unsharp Mask, and Delineate modules for data sharpening and filtration provided by the software. The animated micro-CT scan of the battery can be found in **Video S4-S5**.

## Electrochemical Characterization

### *Cyclic Voltammetry*

The 3-electrode half-cell CV characterization was performed on a cell assembled with the printed electrodes as the working electrode, a platinum foil as the counter electrode, Zn metal foil as the reference electrode, and 2 pieces of KOH-PVA hydrogel as the electrolyte. The 3-electrode full-cell CV characterization was performed between 1.35 V to 2 V on a cell assembled similar to the typical battery architecture but with an extra Zn metal foil as the reference electrode. The structures of both cells are illustrated in **Figure S10**. The CV was performed using an Autolab PGSTAT128N potentiostat/galvanostat with an additional pX-1000 module. In the 3-electrode full-cell CV, the AgO cathode was connected to the working electrode probe, the Zn anode was connected to the counter and reference electrode probes, and the pX-1000 module was used to monitor the potential between the cathode and the reference Zn foil. The potential of the anode vs. Zn was obtained by subtracting the cathode vs. Zn potential from the full cell potential. A scan rate of 10 mV/s was used for all CV tests.

### *Discharging and Cycling Protocol*

The constant current complete discharge of the battery for primary applications was performed with the following procedure. Firstly, the assembled and vacuum-sealed battery was left idle for 1 hr to allow the electrolyte to fully permeate through the electrodes. Then, the battery was discharged using a battery test system (Landt Instruments CT2001A) at the desired current, until reaching the lower cut-off voltage of 1.35 V.

To enable the secondary application of the battery, cycling protocols were established that rely on the accurate control of the potential and DOD of the battery. To perform charge-discharge cycling on a fabricated battery, 50% of its maximum capacity, which was estimated by the low-current complete discharges, was first determined as the cyclable capacity and the basis to determine C-rates of the protocol. The battery was firstly discharged at the C-rate of 0.1C from 100% to 40% DOD. Then, the battery was recharged at the C-rate of 0.2C until reaching 2V, and then at 2V until reaching 90% DOD or C-rate of 0.05C. The battery could be then discharged and recharged at the desired C-rates between 1.35 V and 2 V, with the DOD maintained between 40% and 90% of its maximum value. Unless specified otherwise, all cycling data were performed using cells with 1 × 1 $cm^2$ form factor with 2 layers of active electrode materials. More cycling data for two cells with 8-layer electrode thickness connected in series is shown in **Figure S15**.

The pulsed discharge protocol was designed to simulate the battery's performance in powering a typical MCU-controlled wearable device with integrated BT functionality. The battery was discharged using an Autolab PGSTAT128N potentiostat/galvanostat implementing fast chrono methods. See **Figure S25** for the detailed discharging script.

*Electrochemical Impedance Spectroscopy*

Electrochemical Impedance Spectroscopy (EIS) measurements were performed with a Biologic SP-150 in a 3-electrode configuration. The Zn-AgO three electrodes cell was fabricated with a Zn reference wire placed between an extra layer of hydrogel electrolyte and the original electrolyte layer shown in **Figure 2B**. The Zn reference wire was then connected to an Au sputtered heat-sealable SEBS-based printed carbon tab that was vacuum sealed to ensure complete cell sealing to hinder electrolyte dehydration. The working electrode (WE) and counter electrode (CE) were connected to the AgO cathode and Zn anode, respectively.

The impedances of the two half cells and the full cell were monitored in-situ during charging and discharging to analyze impedance changes most closely related to practical cycling conditions with a galvanostatic-EIS (GEIS) measurement. Accordingly, the DC base current was set to the current of the charging/discharging step, while the AC amplitude was set to 300 µA, approximately one-fifth of the cycling current. The frequency sweep was between 1 MHz and 1 Hz with 10 points per decade and an average of 8 measures per frequency. The cycling script implemented with GEIS is similar to that of the capacity-limited electrochemical cycling protocol, with the exception that the voltage limits applied were 1.95 V and 1.4 V vs. the reference instead of the anode for the charging and discharging respectfully. For each charge and discharge step, 10 GEIS was measured for 15 complete cycles, resulting in a total of 870 separate Nyquist plots (29 steps × 10 measures × 3 cell configurations). For analysis simplicity, only the $5^{th}$ cycle's discharge and charge were analyzed.

Both half-cell Nyquist plots for the 5th cycle's discharge and charge steps were fitted to equivalent circuits using a slightly modified version of the Zfit function available as open-source code from Mathworks.[41] Zfit utilizes another Mathworks open source code, fminsearchbnd, to minimize the error of simulated impedances with the experimental values by altering the impedance parameters (i.e. resistance values, constant phase element values, etc.) under realistic parameter boundary conditions.[42] The use of this code allowed for streamline fitting of many successive Nyquist to provide insights in observable trends in the fitted parameters. Additional data of the EIS measurement can be found in **Figure S16-18**.

*Electrolyte Conductivity Measurement*

The ionic conductivity of the gel electrolyte was measured by a customized two-electrode (Stainless Steel 316L) conductivity cell with an inner impedance at 0.54 Ω. The cell constant is frequently calibrated by using OAKTON standard conductivity solutions at 0.447, 1.5, 15, and 80 $mS\cdot cm^{-1}$ respectively. A constant thickness spacer was positioned between the two electrodes which ensure no distance changes during multiple-time measurements. The electrolytic conductivity value was obtained with a floating AC signal at a frequency determined by the phase angle minima given by Electrochemical Impedance Spectroscopy (EIS) using the following equation: $\sigma = KR^{-Q}$, where R is the tested impedance (Ω), K is the cell constant ($cm^{-1}$) and Q is the fitting parameter.[43] All of data acquisition and output were done by LabView Software, which was also used to control an ESPEC BTX-475 programming temperature chamber to maintain the cell at a set temperature in 30 minutes intervals.

<u>Mechanical Deformation Tests</u>

The bending deformation of the battery was conducted by bending a 1 × 5 $cm^2$ battery around a cylinder with the diameter of 1 cm manually. The deformation was cycled between the bent and relaxed state at the rate of 1 s/cycle for 100 cycles. See **Figure S19** for the bending control of the battery mechanical tests. Similarly, the twisting deformation of the battery was performed manually at 1 s/cycle by fixing one end of the battery and twisting the other end 180° clockwise and counterclockwise for 100 cycles.

Assembly of Flexible Display Electronics

To demonstrate the battery's ability to power flexible electronics, a Waveshare 2.9-inch e-Paper flexible display was powered by two Zn-AgO batteries in series. The display module was connected to an Adafruit Feather nRF52 Bluefruit Low Energy (LE) chip and programmed using Arduino and C. The picture of the assembled system is shown in **Figure S23**. MATLAB code was used to convert images to hexadecimal format to be uploaded to the board and the display. The BluefruitConnect IOS app was used to connect the Adafruit chip via Bluetooth to change the display. The system diagram of the E-ink display system is shown in **Figure S22**. The pulsed current profile needed to power the Bluetooth chip and display was determined using an oscilloscope by measuring the voltage across a 10 Ω resistor connected in series with the circuitry. A model pulsed profile was then extracted to be applied to flexible batteries for further testing.

## Acknowledgments

This work was supported by funding from ZPower LLC and Qualcomm. This work was performed in part at the San Diego Nanotechnology Infrastructure (SDNI) of UCSD, NANO3, a member of the National Nanotechnology Coordinated Infrastructure, which is supported by the National Science Foundation (Grant ECCS-1542148). The authors would also like to acknowledge the National Center for Microscopy and Imaging Research (NCMIR) technologies and instrumentation are supported by grant R24GM137200 from the National Institute of General Medical Sciences. AgO, Zn, and $TiO_2$ used in this work were provided by ZPower LLC. SEBS used in this work was provided by Kraton.

## Conflict of Interest

The authors declare no conflict of interest.

**TOC entry**

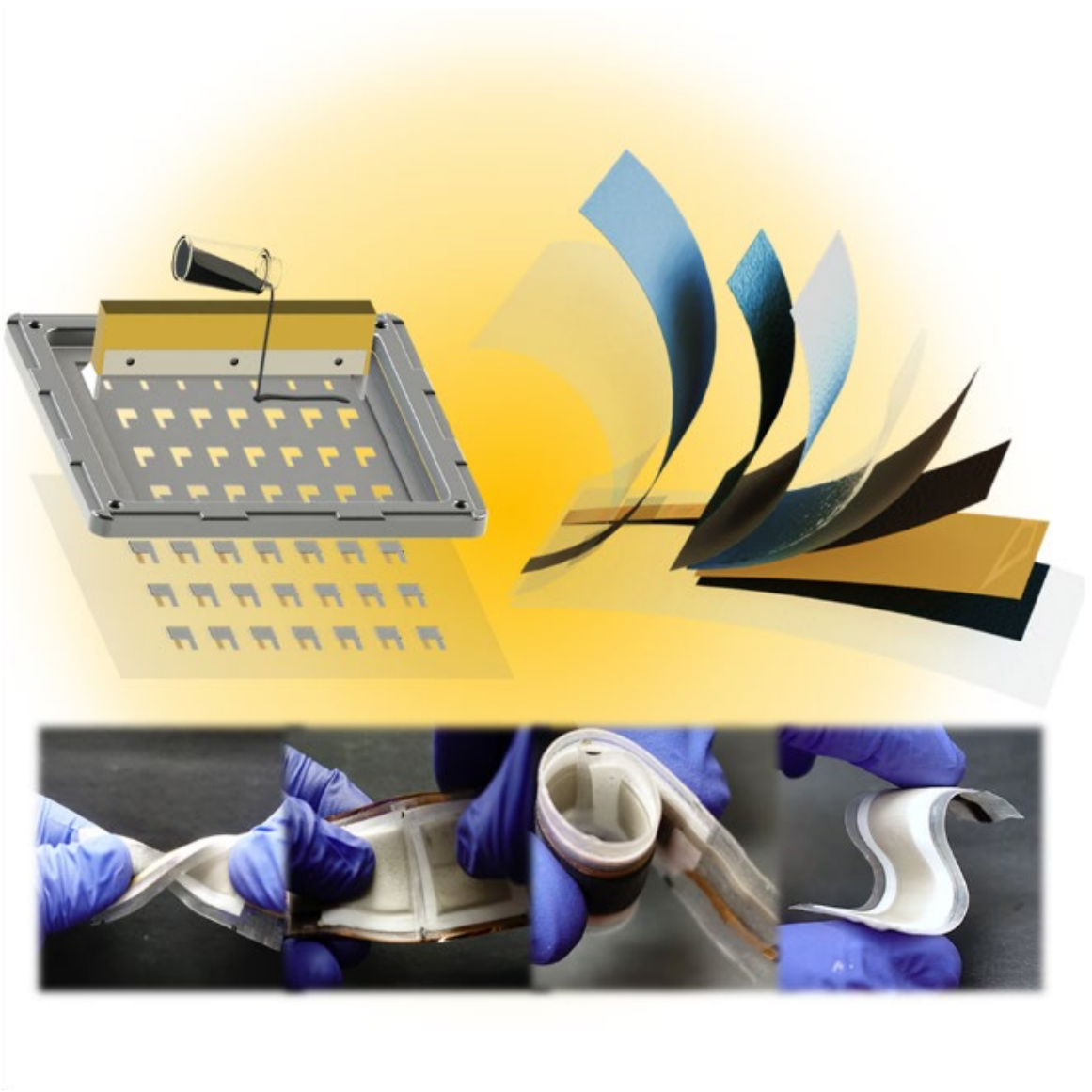

Highlights

Novel polymer-based printing fabrication of battery with a high areal density of 54 $mAh/cm^2$.

The battery is flexible, rechargeable, low impedance, customizable, and low-device footprint.

Superior battery performance in pulsed high current discharge mode.

# Supplementary Information

# High Performance Printed AgO-Zn Rechargeable Battery for Flexible Electronics

Lu Yin, Jonathan Scharf, Jessica Ma, Jean-Marie Doux, Christopher Redquest, Viet Le, Yin Yijie, Jeff Ortega, Xia Wei, Joseph Wang, and Ying S. Meng

## Content

## Supplementary Tables

**Supplementary Table 1**. Comparison of areal capacities of various published and commercialized thin-film batteries. The summary is also visualized in **Figure S1**.

| **Battery Type** | **Fabrication** | **Max. Areal Capacity ($mAh/cm^2$)** | **Configuration** | **Cycle Number** | **Ref** |
|---|---|---|---|---|---|
| $Zn-Ag_2O$ | Screen Printing | 1.6 | In-plane | 11 | [1] |
| $Zn-Ag_2O$ | Screen Printing | ~3 | In-plane | 30 | [2] |
| $Zn-Ag_2O$ | Extrusion Printing | ~2.8 | In-plane | Primary | [3] |
| $Zn-Ag_2O$ | Screen Printing | 1.5 | In-plane | 13 | [4] |
| $Zn-MnO_2$ | Plotting | ~2.2 | In-plane | 30 | [5] |
| $Zn-MnO_2$ | Doctor Blade | <0.077 | Stacked | 140 | [6] |
| $Zn-Ag_2O$ | Drop-cast/ Electroplating | 0.11 | In-Plane | 33 | [7] |
| $Zn-MnO_2$ | Screen Printing | 5.6 | Stacked | Primary | [8] |
| $Zn-Ag_2O$ | Screen Printing | 5.4 | Stacked | Primary | [9] |
| Zn-Ag2O | 3D Printing | 2.4 | Interdigitated Pillars | 7 | [10] |
| $Zn-MnO_2$ | Soaking | 3.775 | In-plane | Primary | [11] |
| LTO-LFP | 3D Printing | 1.5 | In plane Interdigitated | 30 | [12] |
| Zn-Air | Screen Printing | 1.4 | Stacked | Primary | [13] |
| $Zn-Ag_2O$ | Screen Printing | 11 | Stacked | Primary | [14] |
| Si/CNT-NMC | Doctor Blade | 30 | Stacked | 50 | [15] |
| $Zn-MnO_2$ | NA | ~1 | In-plane | Primary | [16] |
| $Zn-MnO_2$ | NA | ~3 | NA | Primary | [17] |
| $Zn-MnO_2$ | NA | ~1.5 | NA | Primary | [18] |
| $Li-MnO_2$ | NA | ~2.5 | Stacked | Primary | [19] |
| Zn-AgO | Screen Printing | 54 | Stacked | 80 | This work |

**Supplementary Table 2**. KOH-PVA electrolyte information

| Caustic Concentration | Removed Precursor Water wt% | $\sigma_0$ (mS/cm) | $E_a$ (eV) |
|---|---|---|---|
| 26.3% | 65.77% | $2.037\times10^4$ | 0.109 |
| 31.8% | 70.72% | $3.155\times10^4$ | 0.115 |
| 36.5% | 73.88% | $6.029\times10^4$ | 0.138 |

## Supplementary Figures

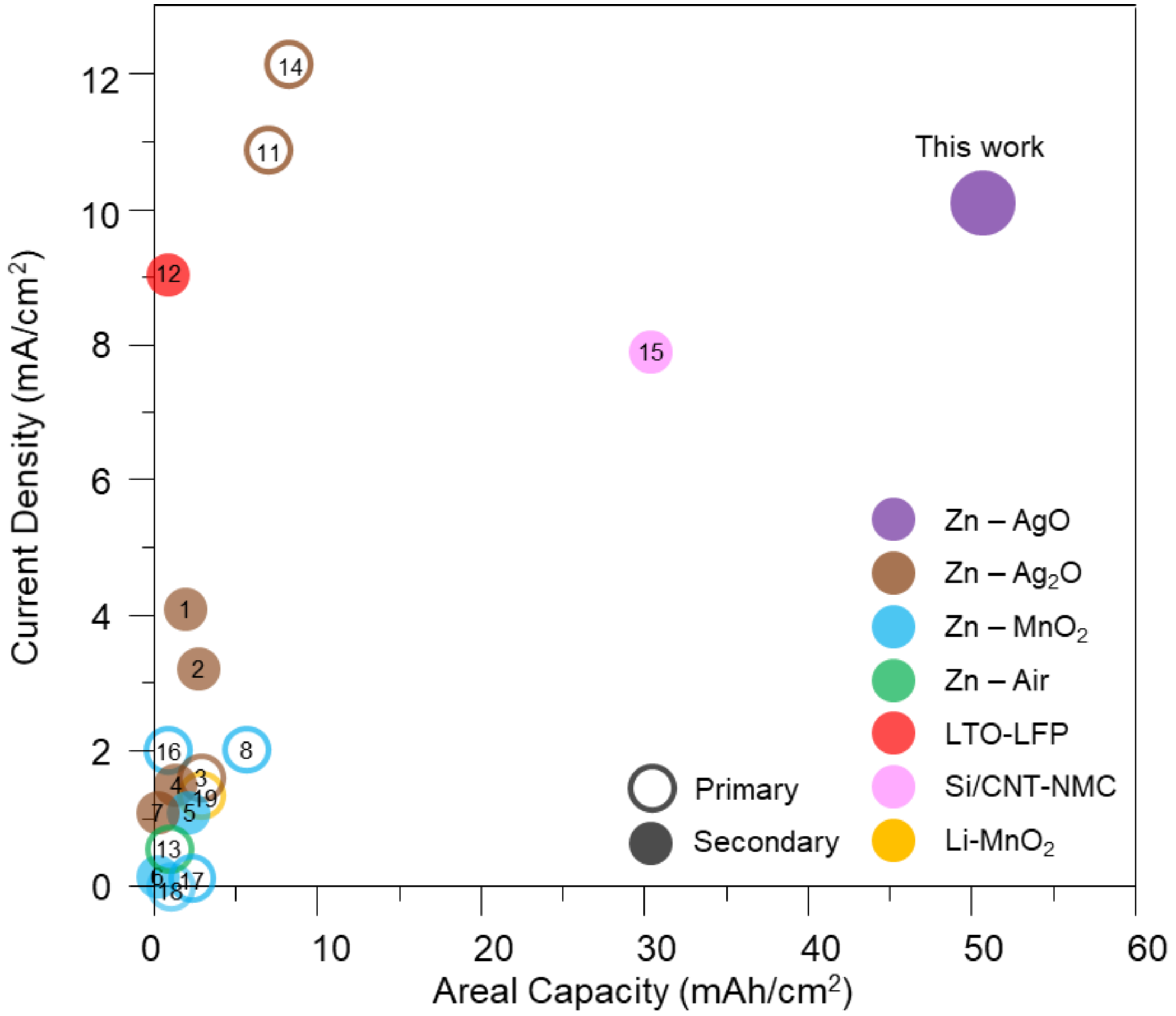

**Supplementary Figure 1**. Visualization of **Table S1** comparing the maximum current densities and areal capacities of various printed batteries.

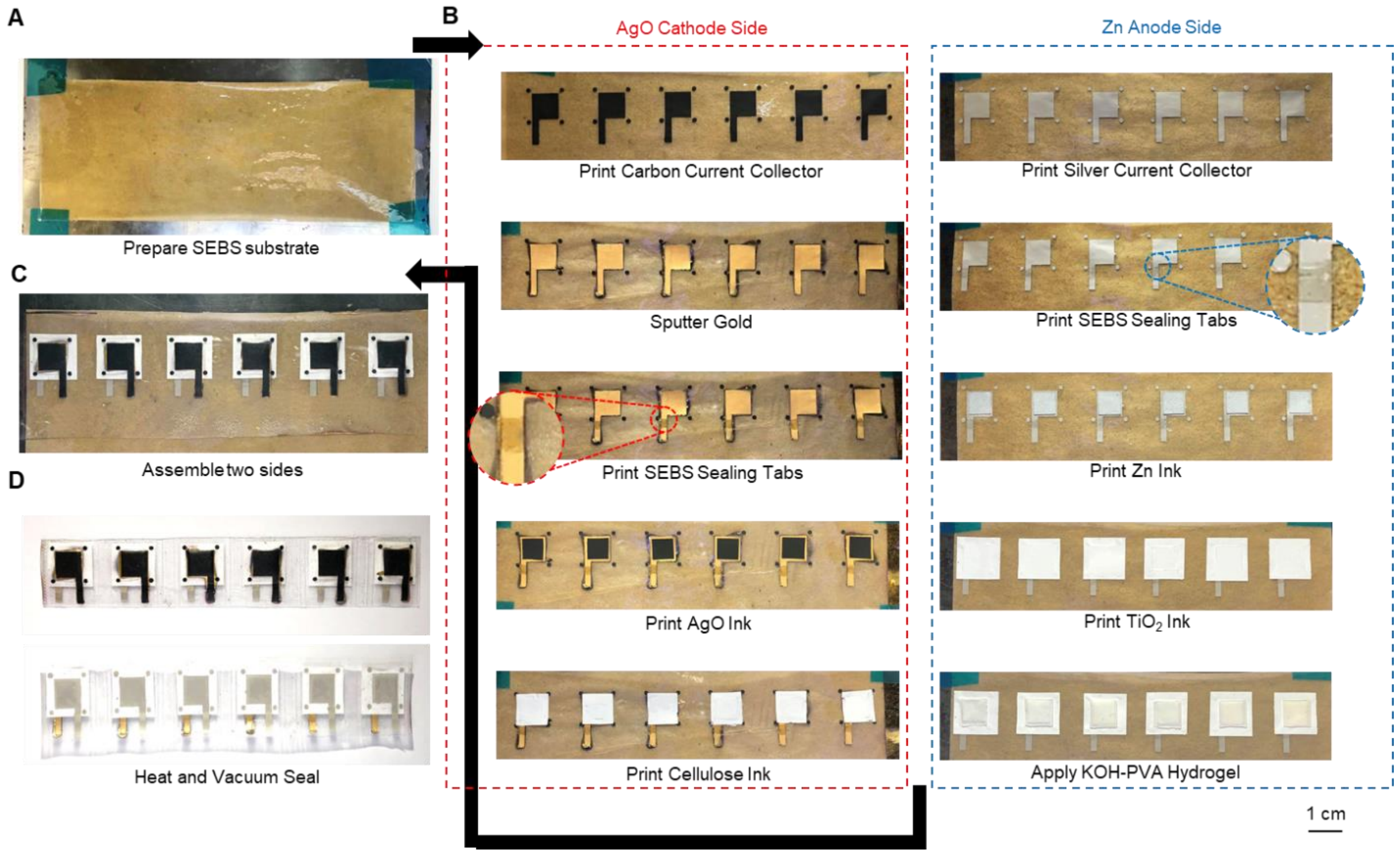

**Supplementary Figure 2**. The images of the step-by-step batched fabrication of the printed AgO-Zn batteries. (A) Prepared SEBS substrate. (B) The layer-by-layer printing of the AgO cathode (left) and the Zn anode (right). (C) Placing the cathode side onto the anode side with the hydrogel electrolyte in between. (D) Heat and vacuum sealing of the batteries. Each cell was separated by further heat sealing after the entire batch was vacuum sealed.

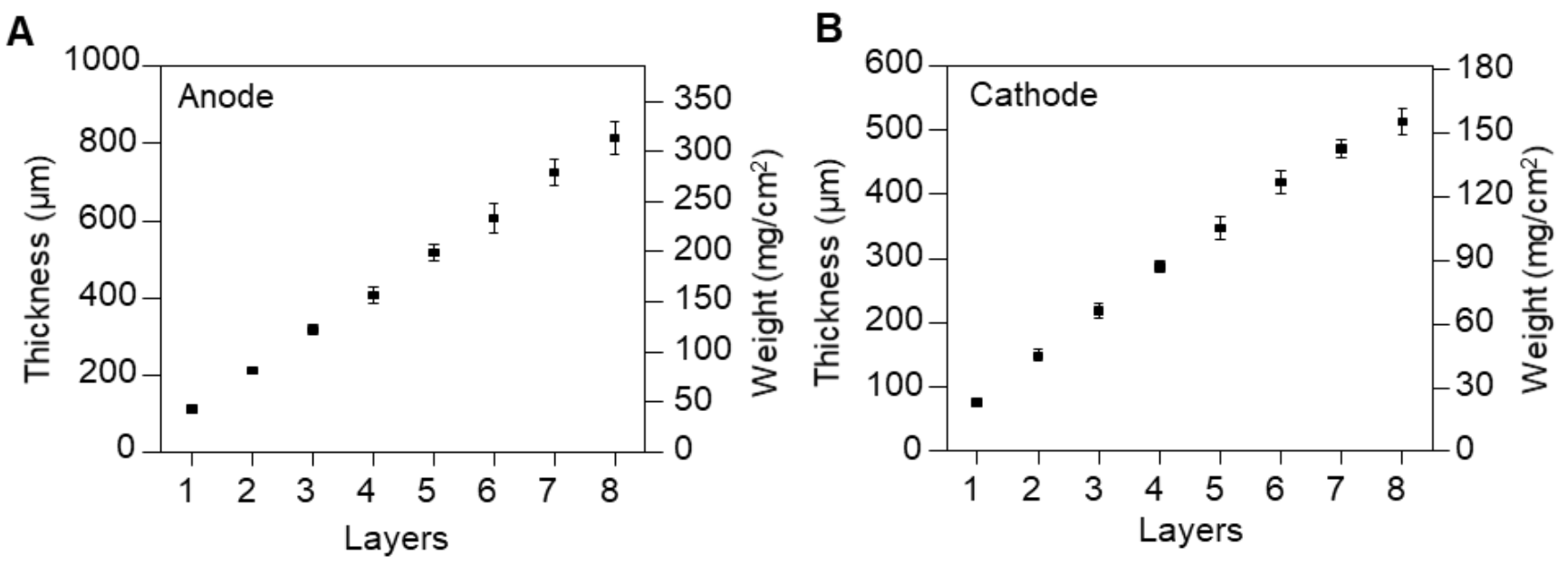

**Supplementary Figure 3**. Thickness calibration of the (A) anode and (B) cathode printed using their corresponding stencils. 5 samples were taken to generate the average thicknesses and standard deviations of each data point.

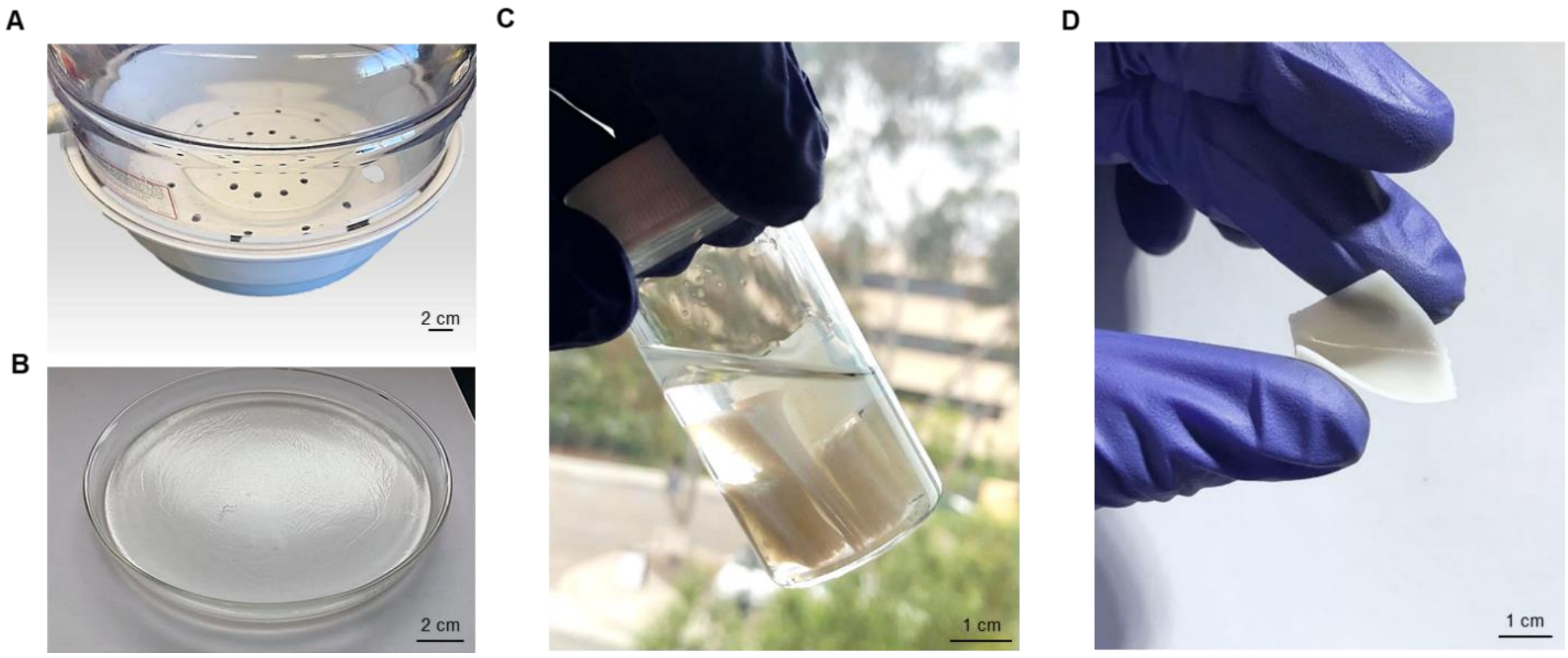

**Supplementary Figure 4**. Images of the fabrication of the KOH-PVA electrolyte gel. (A) Drying of the precursor solution to desired concentration in a vacuum desiccator. (B) The crosslinked 36.5 % hydrogel after drying. (C) Storage of the hydrogel pieces after cutting into desired sizes. (D) A bent 2 × 2 $cm^2$ hydrogel.

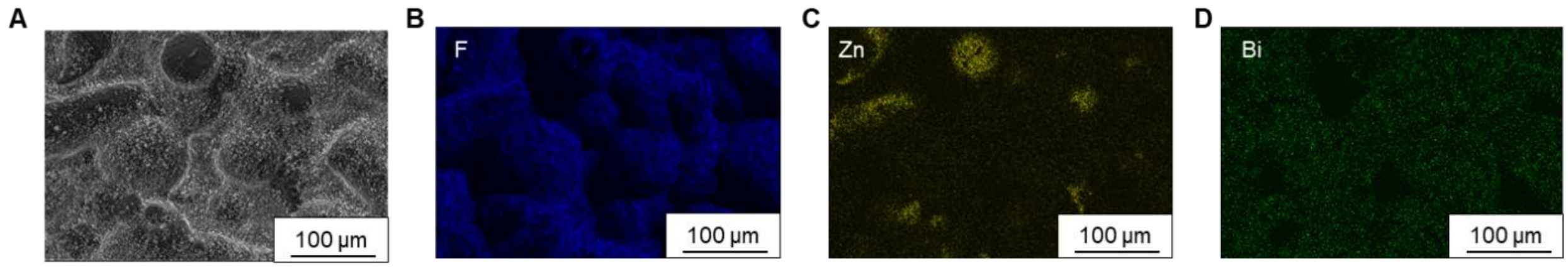

**Supplementary Figure 5**. (A) The SEM and corresponding EDX mapping of (B) fluorine (from the binder), (C) Zn, and (D) bismuth of the anode.

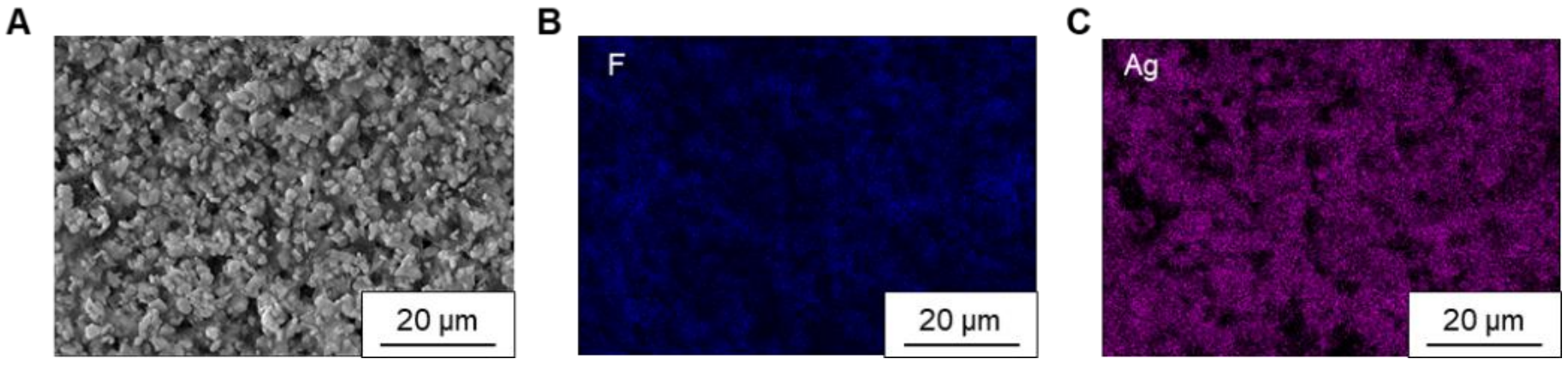

**Supplementary Figure 6** (A) The SEM and corresponding EDX mapping of (B) fluorine (from the binder) and (C) Ag of the cathode.

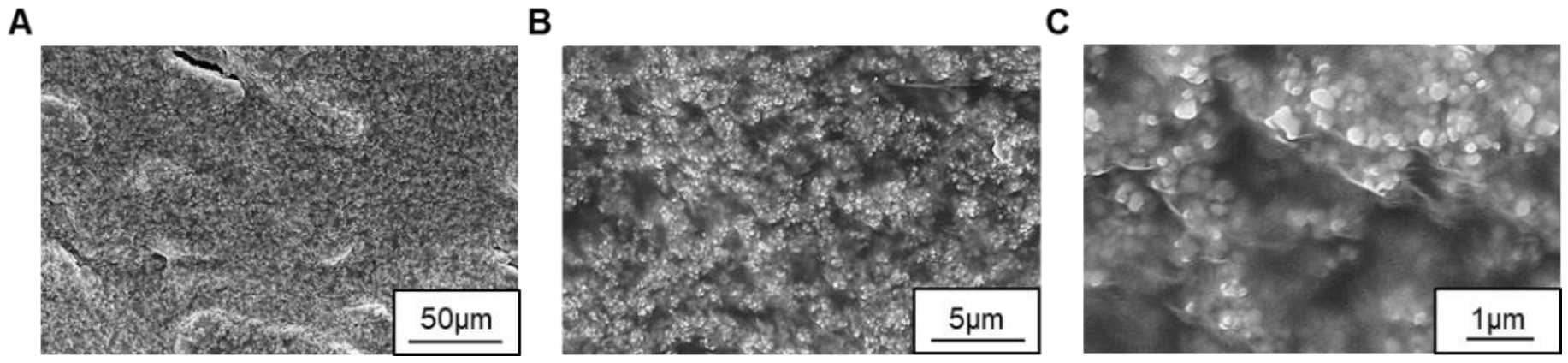

**Supplementary Figure 7**. Additional SEM images of the printed $TiO_2$ separator with different magnifications.

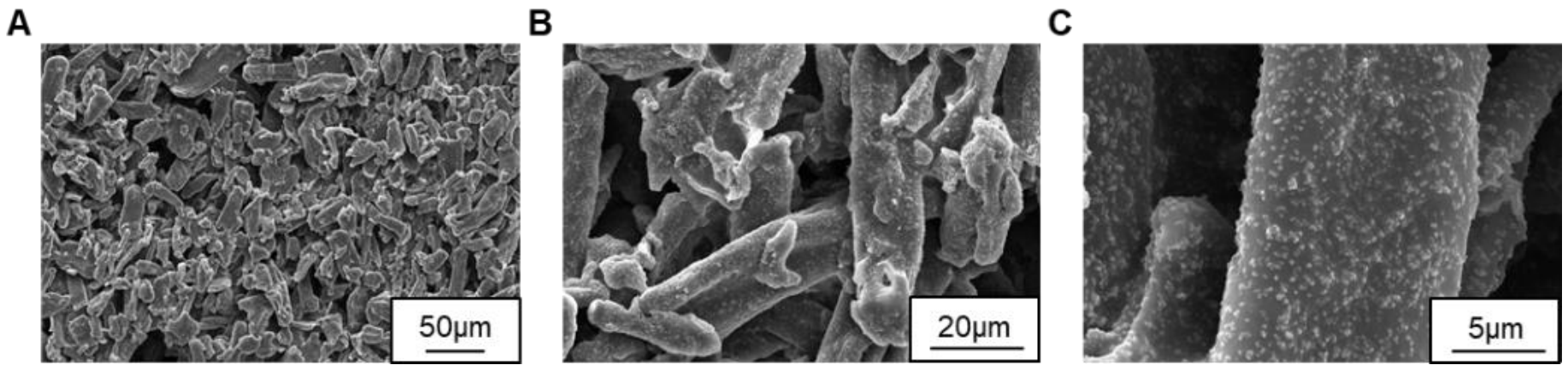

Supplementary Figure 8. Additional SEM images of the printed cellulose separator with different magnifications.

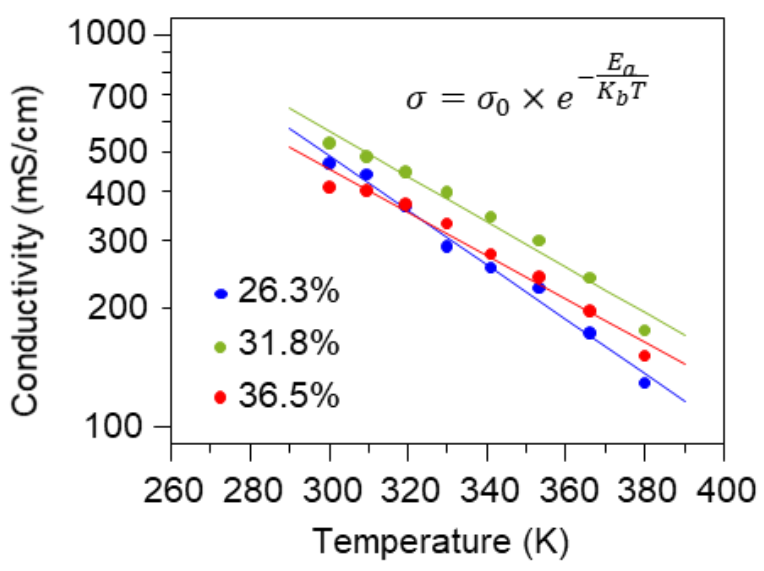

**Supplementary Figure 9**. The conductivity of the hydrogel with different caustic material concentrations. The linear trendline was fitted using the given equation and listed in **Table S2**.

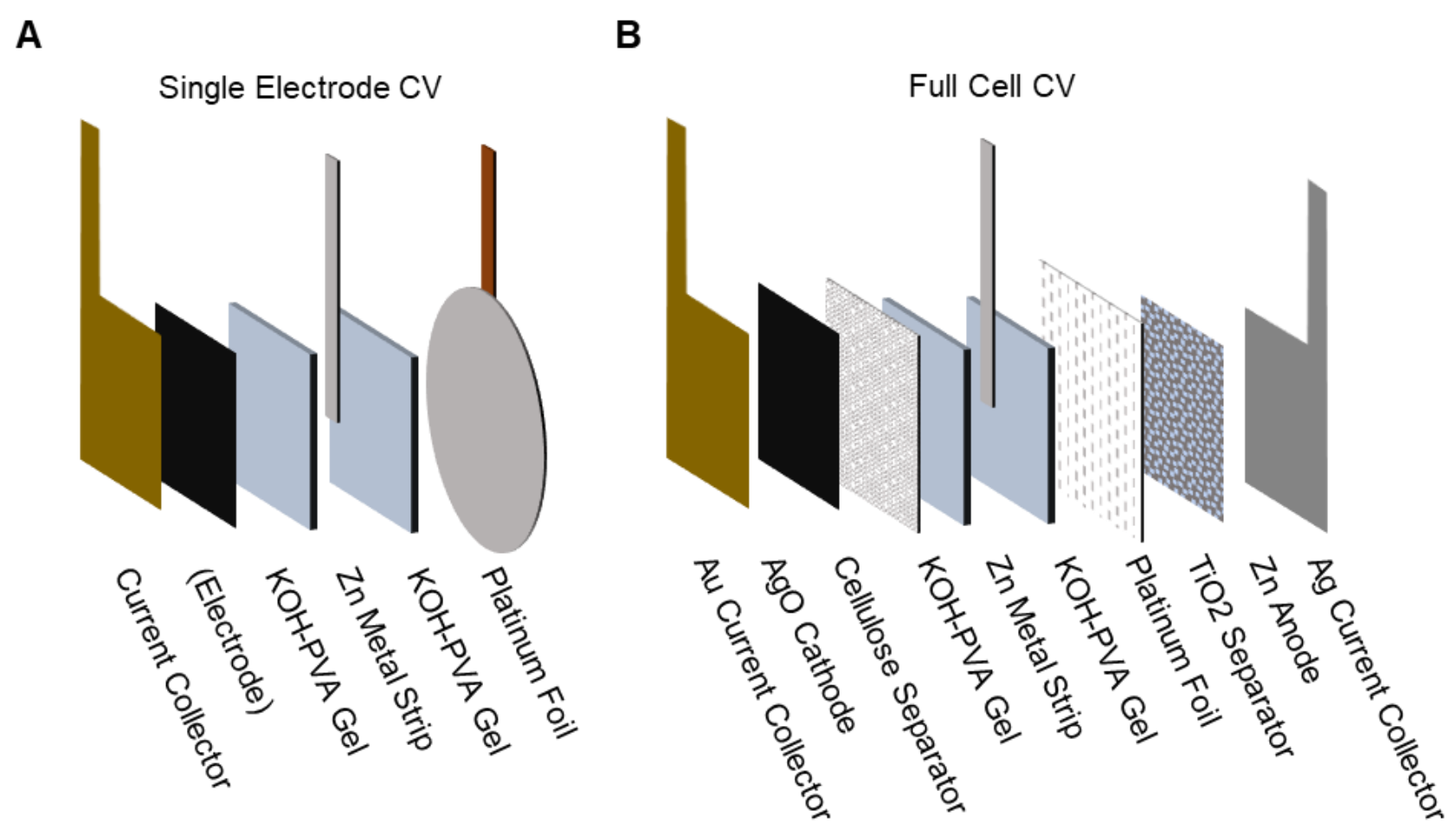

**Supplementary Figure 10**. The cell structure used for the CV analysis. (A) The cell structure used for single electrode scanning for testing the current collectors. (B) The cell structure used for full cell scanning with an external Zn metal strip as the reference electrode.

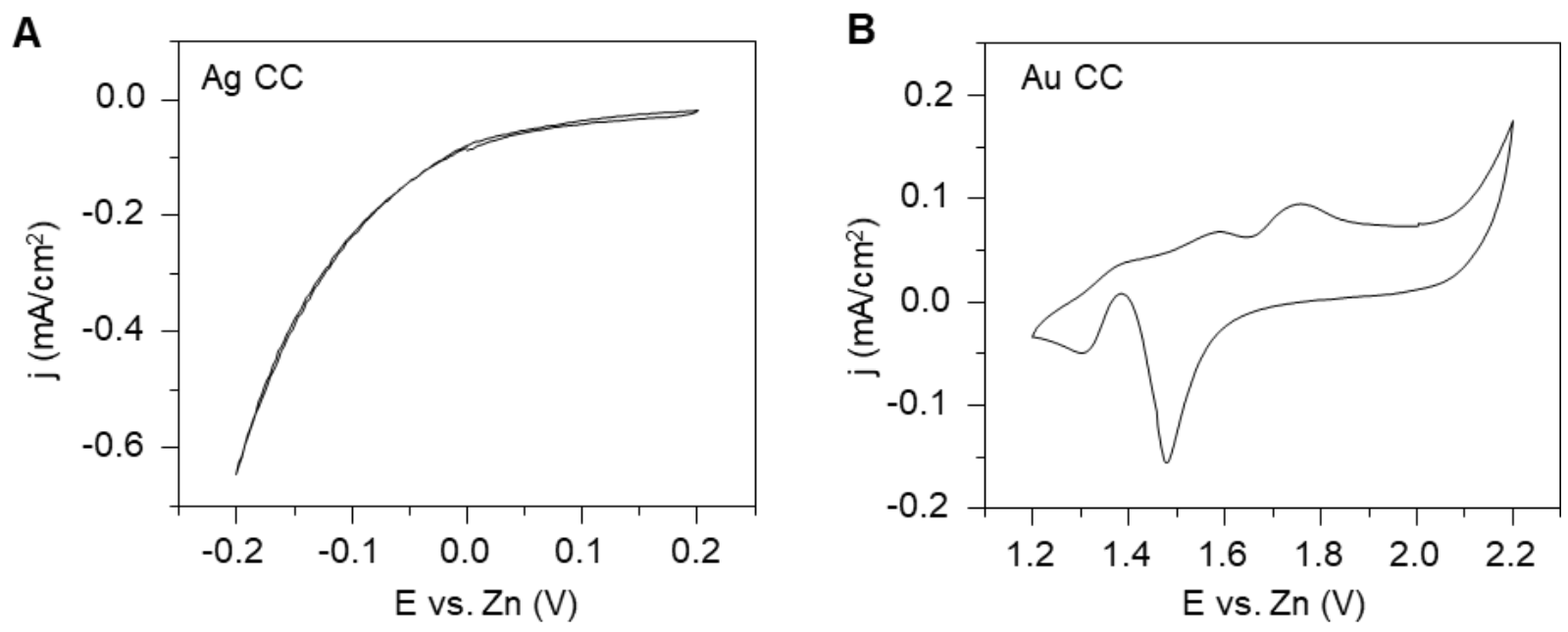

**Supplementary Figure 11**. The CV of the printed (A) Ag anode current collector (CC) and the Au-sputtered carbon cathode CC in their corresponding voltage range used in **Figure 2D-ii**. Scan rate: 10 mV/s.

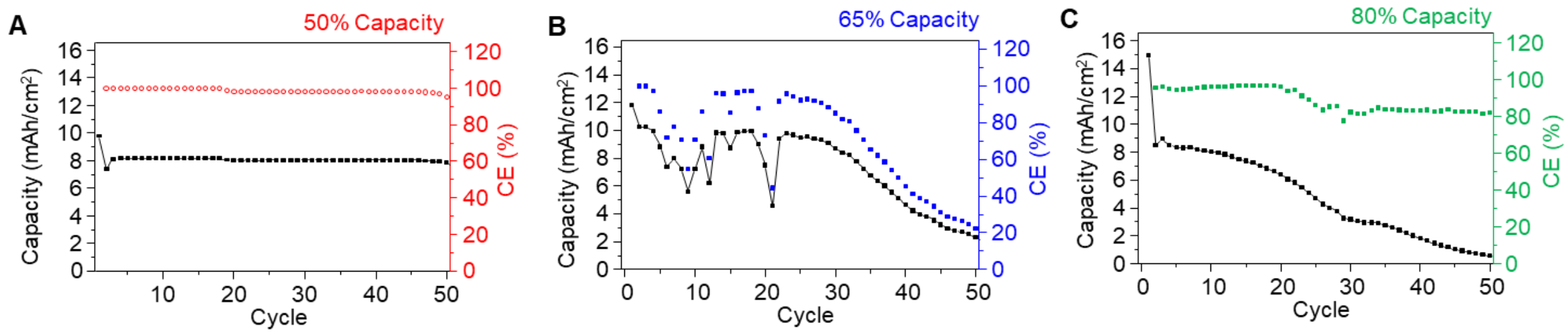

**Supplementary Figure 12.** The cycling of the battery at the different capacity range. (A) Cycling the battery between 40 % and 9 0% state of charge (50 %). (B) Cycling the battery between 2 5% and 90 % state of charge (65 %). (C) Cycling the battery between 10 % and 90 % state of charge (80 %). Electrolyte with the concentration of 36.5 % was used, and the cells were cycled at the rate of 0.2 C.

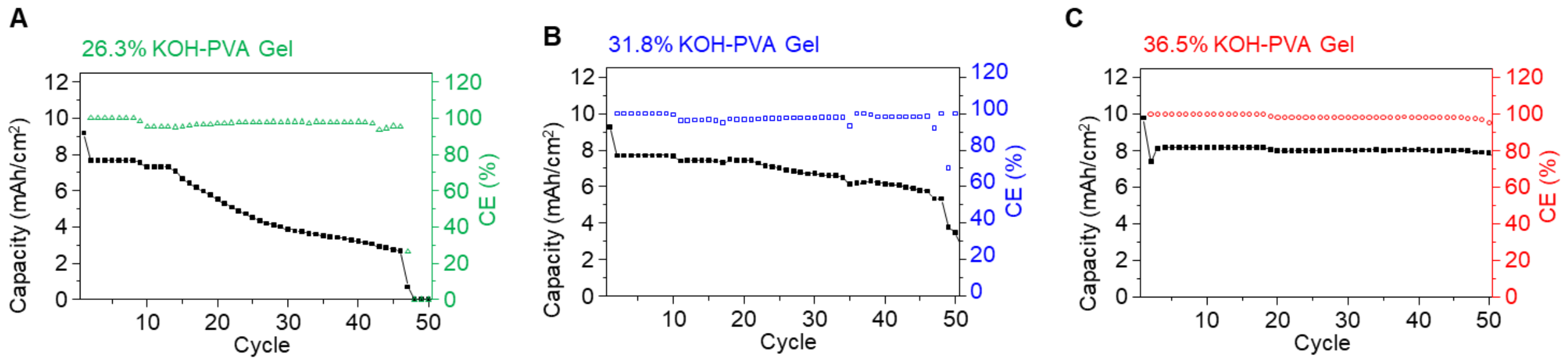

**Supplementary Figure 13**. The cycling of the battery with electrolyte concentration of (A) 26.3 %, (B) 31.8 %, and (C) 36.5 %. The 50 % capacity range was used and the cells were cycled at the rate of 0.2 C.

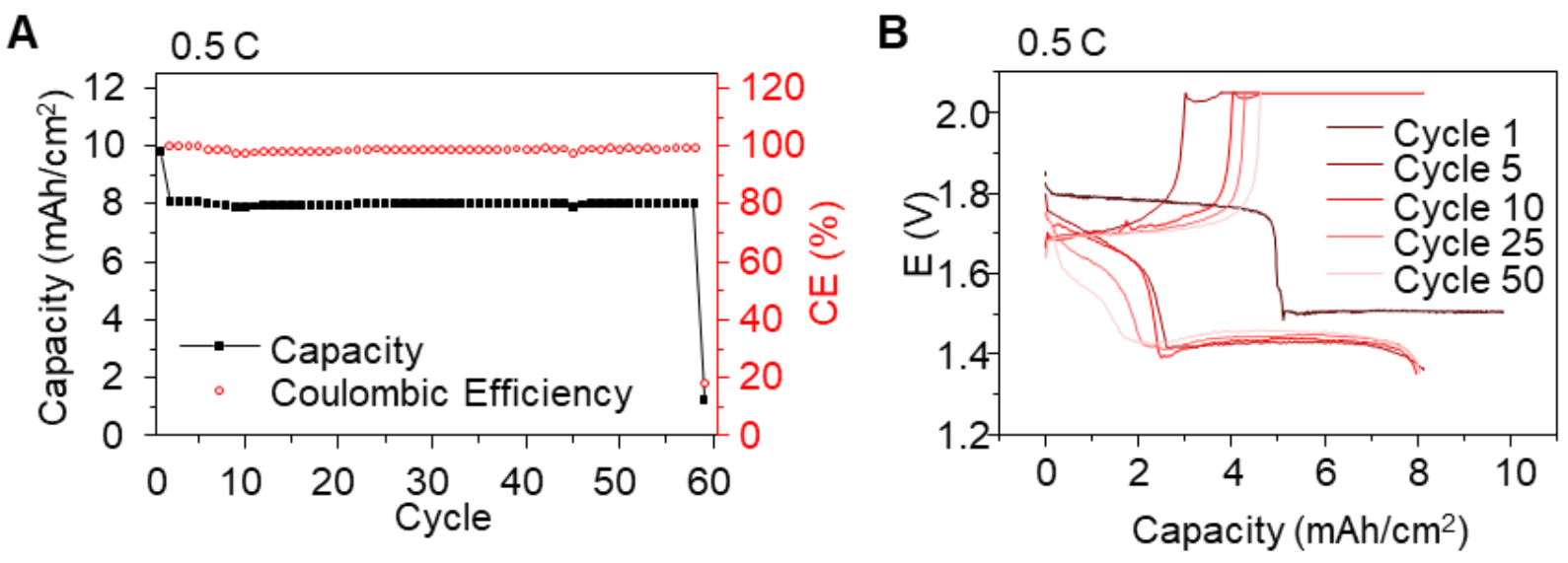

**Supplementary Figure 14**. The cycling of the battery at the rate of 0.5 C. The electrolyte with the concentration of 36.5% and the capacity range of 50% was used.

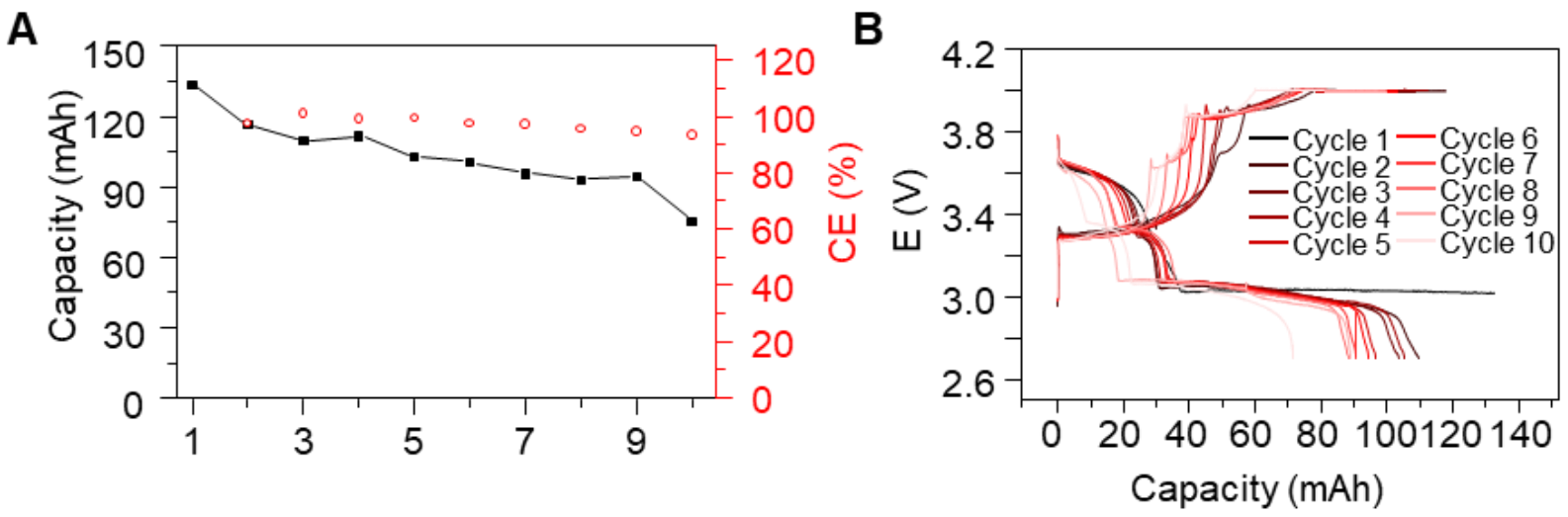

**Supplementary Figure 15**. The cycling of two 8-layer 2 × 2 $cm^2$ battery connected in series at the rate of 0.05 C.

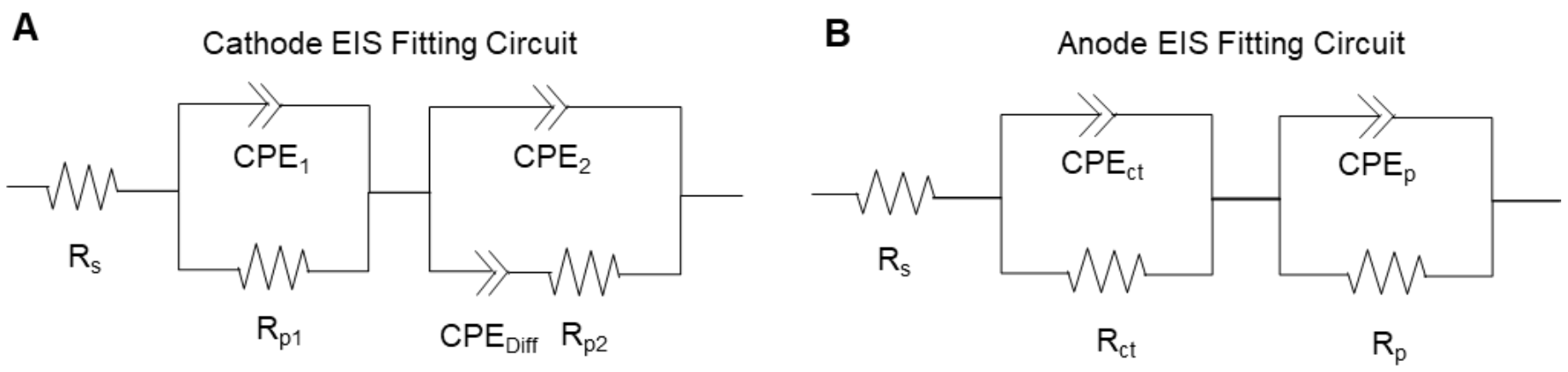

**Supplementary Figure 16**. The equivalent circuit used for the (A) cathode and (B) anode EIS fitting.

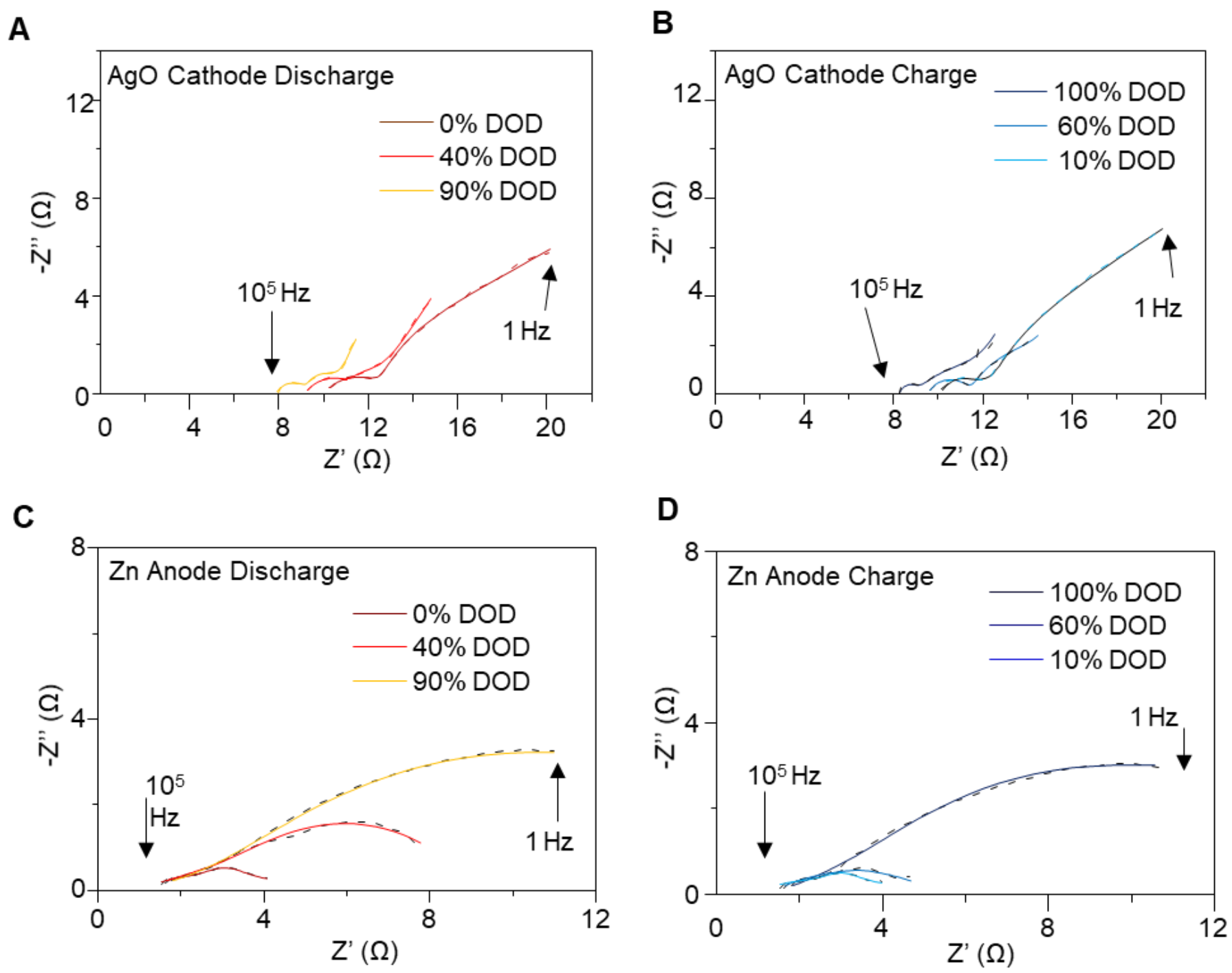

**Supplementary Figure 17**. The Nyquist plot and the EIS fitting of the cathode during the 5$^{th}$ cycle (A) discharging and (B) charging, and the corresponding anode (C) charging and (D) discharging.

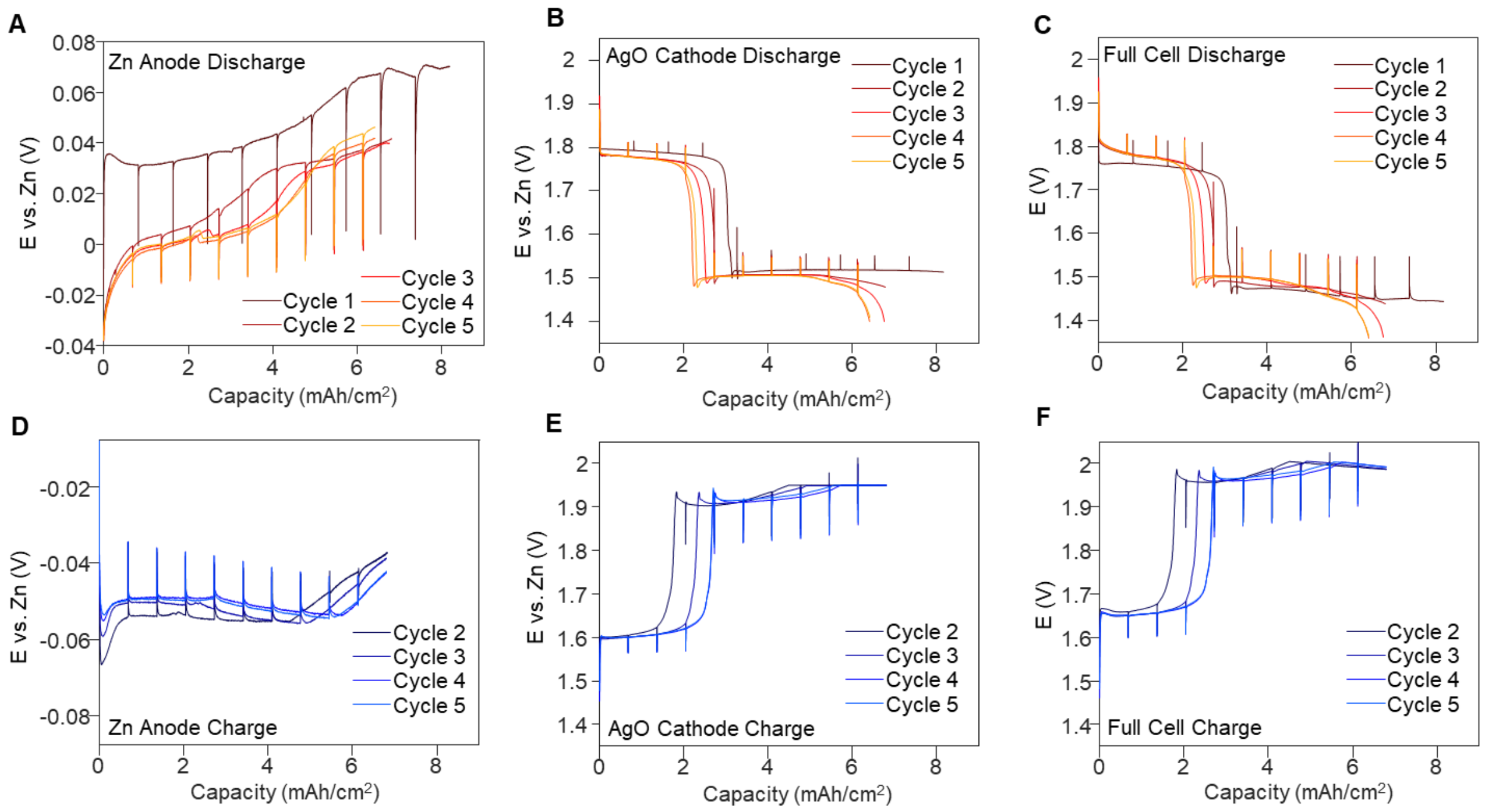

**Supplementary Figure 18**. The potential profile of the anode (A,D) and cathode (B,E) vs. Zn reference and the full cell (C, F) within the first 5 cycles of (A-D) discharging and (D-F) corresponding 4 cycles of charging. The vertical lines correspond to the places where an EIS measurements were taken.

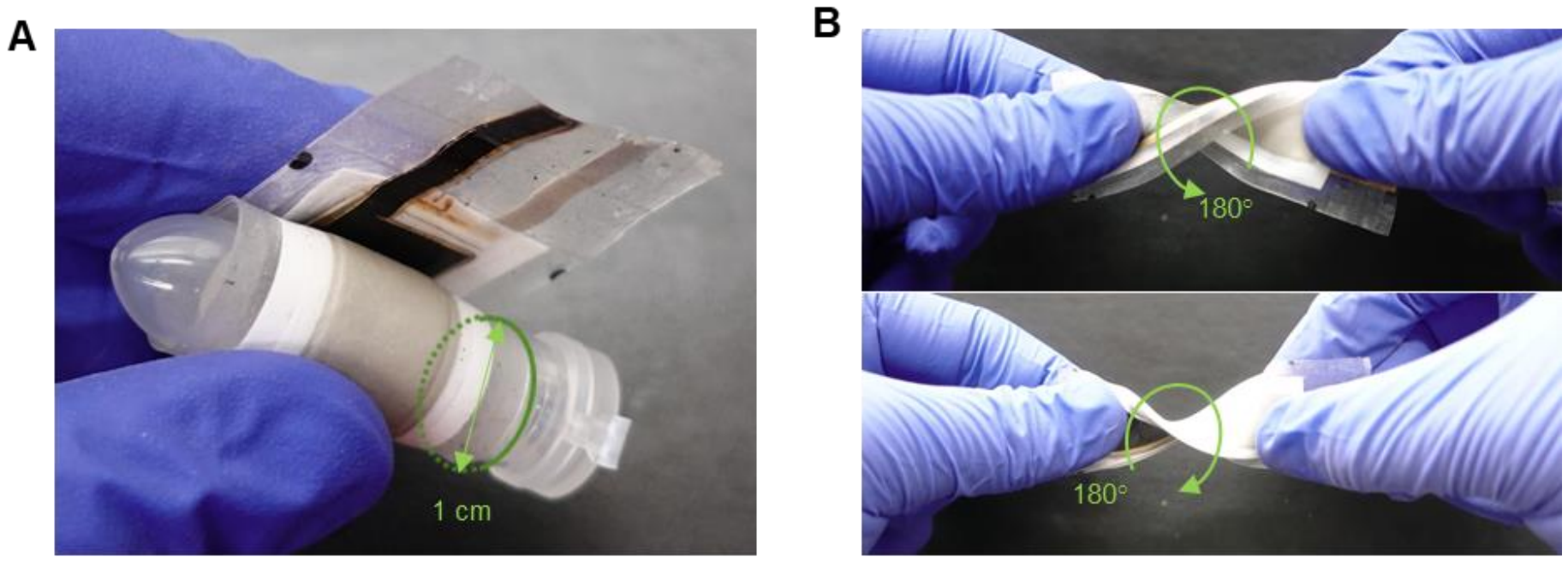

**Supplementary Figure 19**. Additional images illustrating the manual bending and twisting of the battery. (A) A tube with diameter of 1 cm was used to bend the battery for half and one entire round. (B) The battery was twisted counterclockwise and clockwise 180° which add up to a total of 360°.

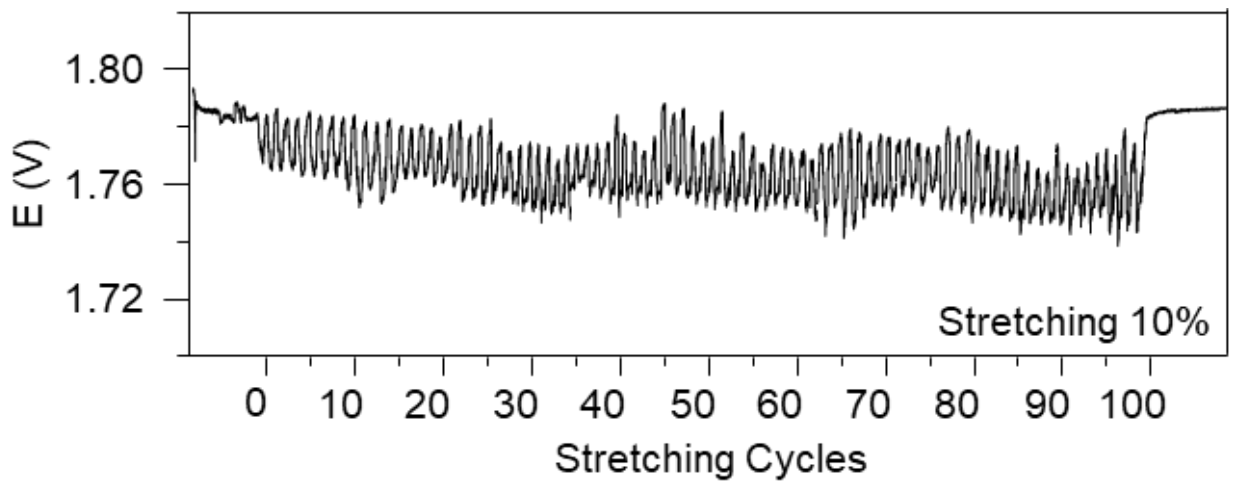

**Supplementary Figure 20**. The voltage profile of the 1 × 5 $cm^2$ battery during 1 mA discharge while undergoing 100 cycles of 10 % lengthwise stretching. Although the battery was not optimized for stretchability, a certain amount of stretchability is required for the battery to endure low-radius bending and accommodate for the outer-layer strain.

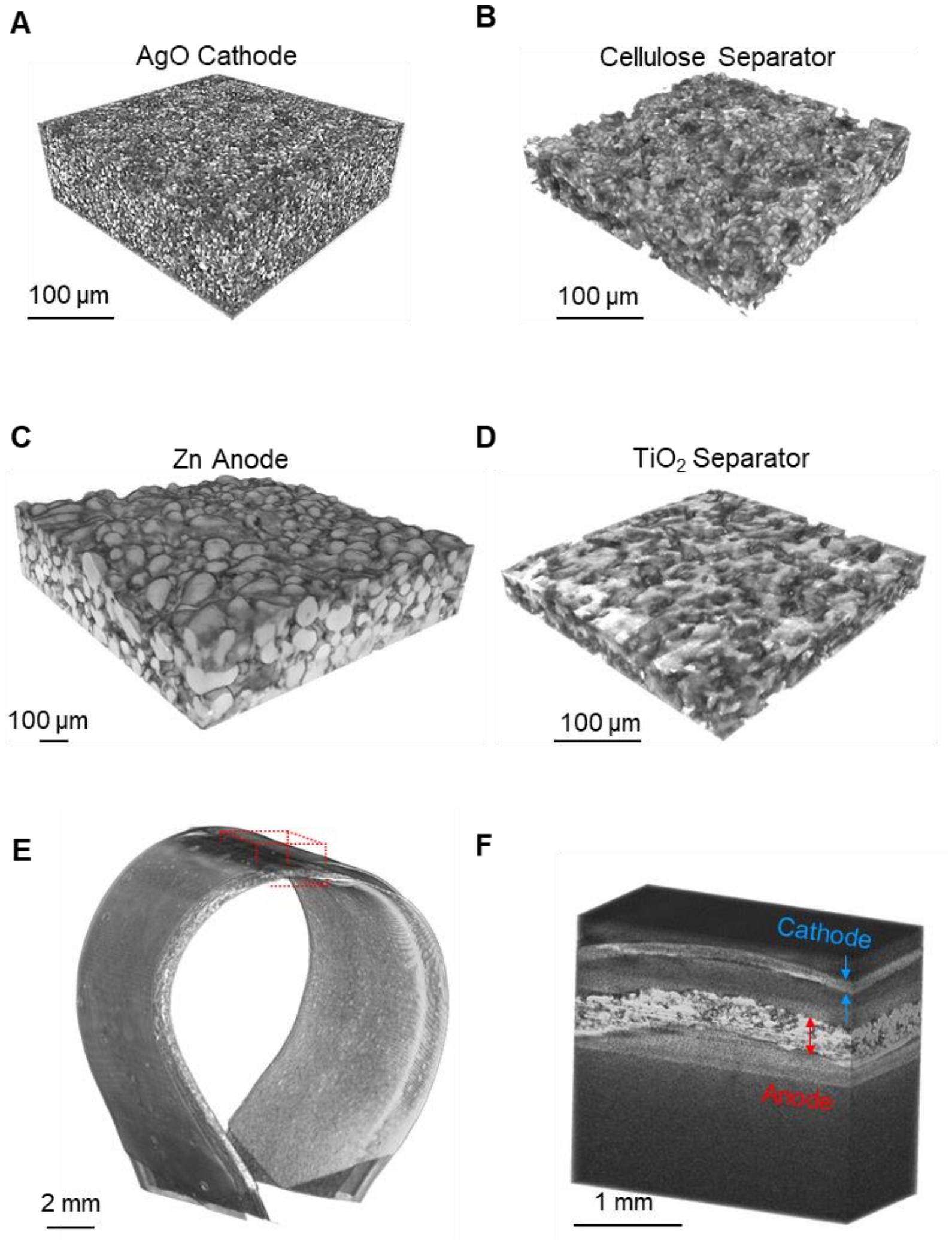

**Supplementary Figure 21**. Additional microscopic 3D images of the (A) cathode, (B) cellulose separator, (C) anode, and (D) $TiO_2$ separator generated using the micro-CT. (E) the 3D image of the bent 1 × 5 cm$^2$ battery in a different angle and (F) the zoomed-in view of the top of the cell showing no cracking nor delamination between the layers.

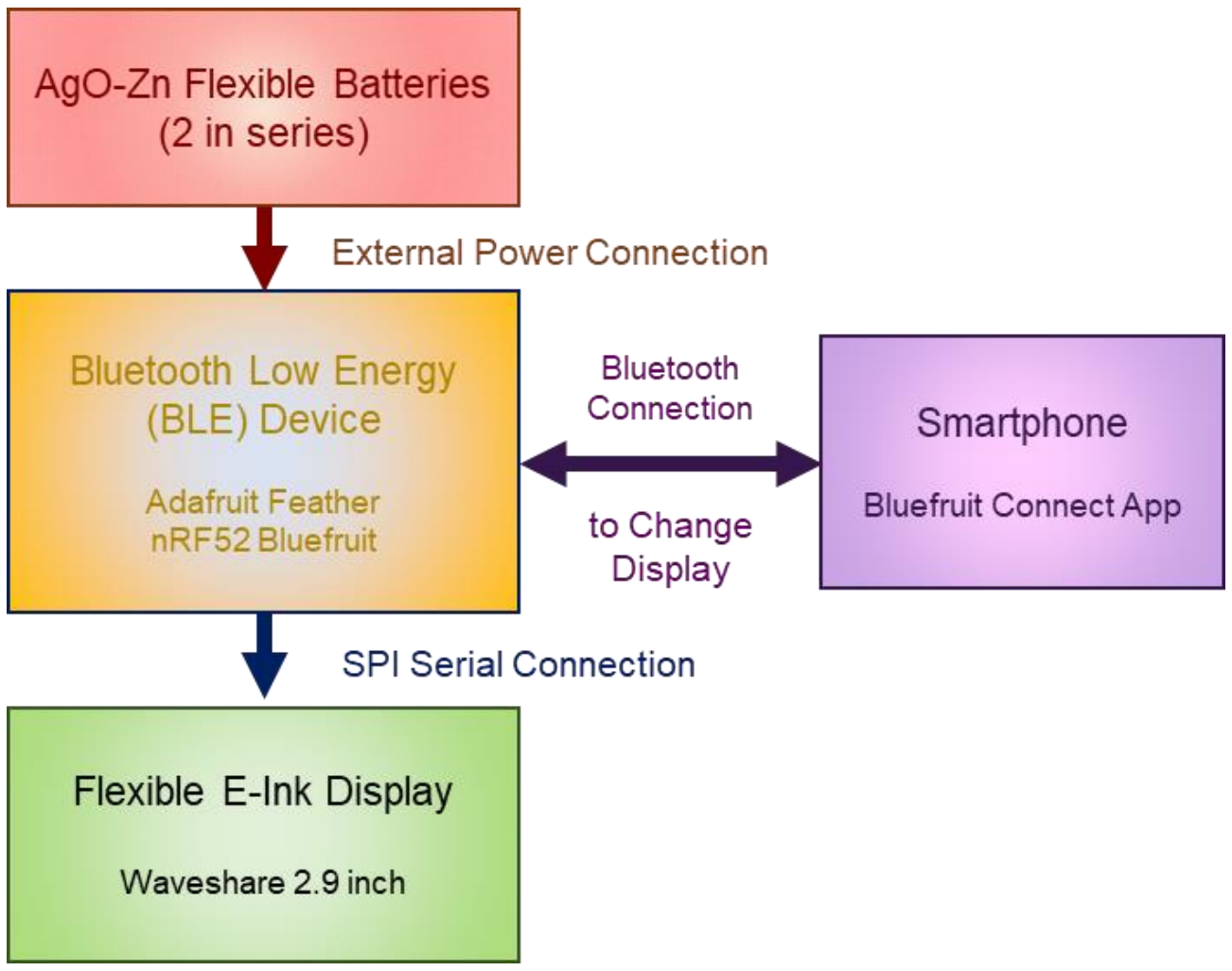

**Supplementary Figure 22**.The system diagram of the assembled flexible E-ink display system.

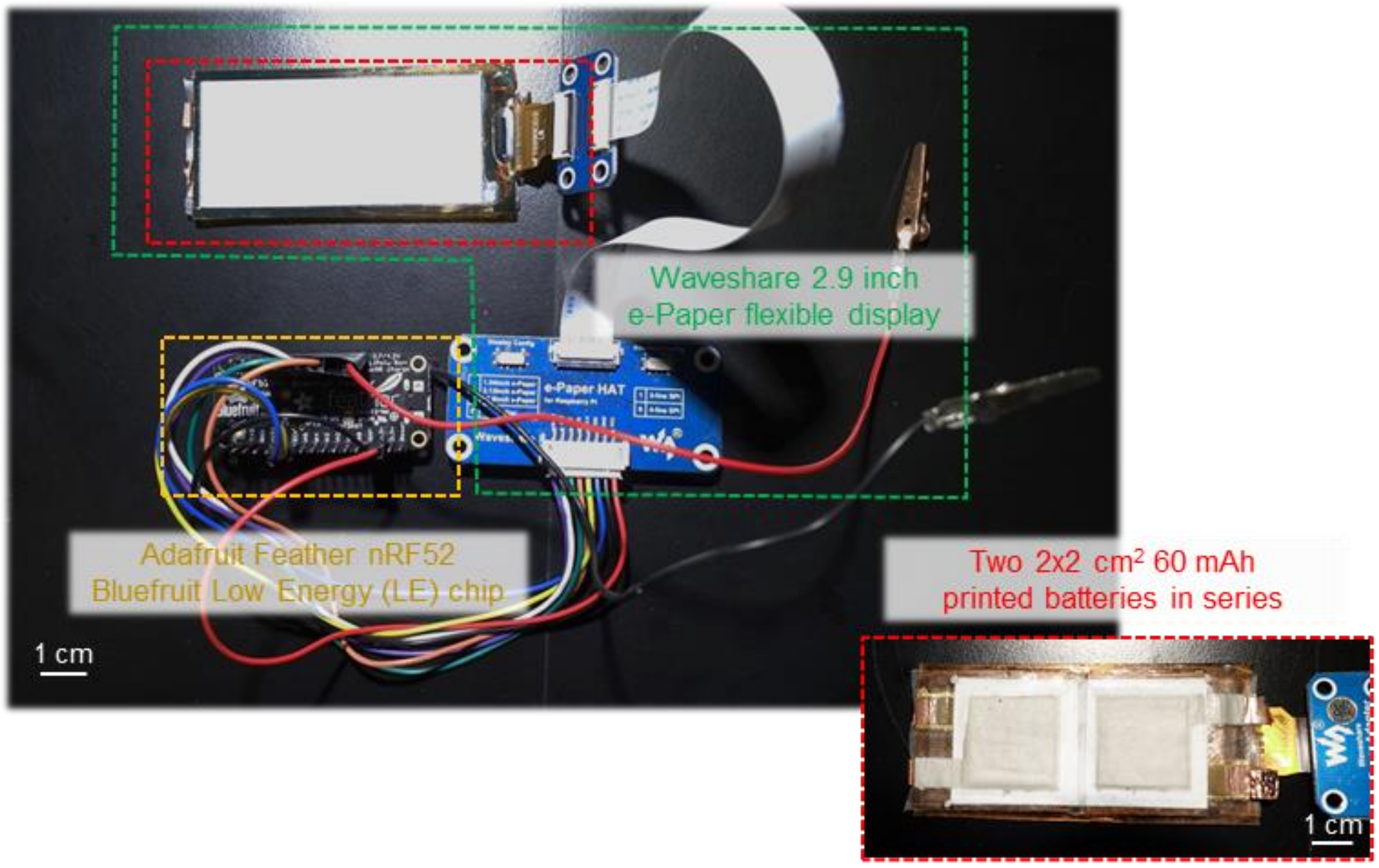

**Supplementary Figure 23**. The photo of the assembled flexible E-ink display system with 2 batteries attached to the backside of the display panel.

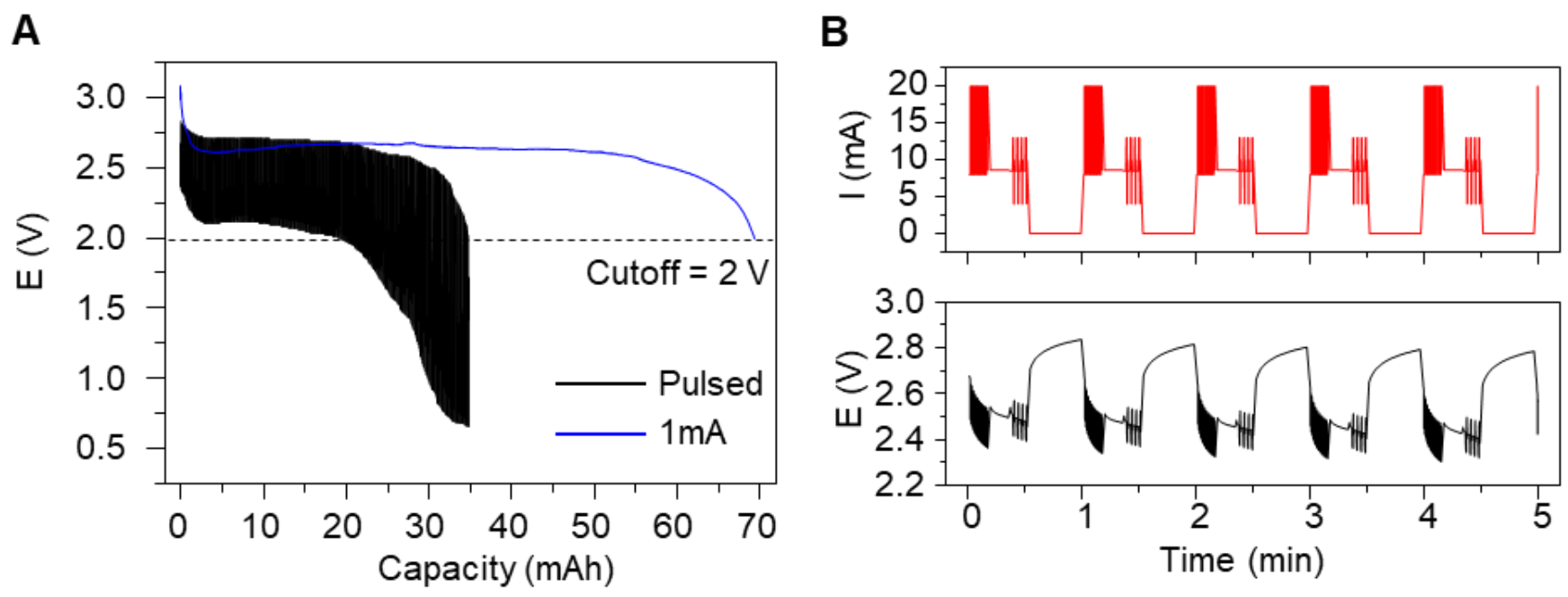

**Supplementary Figure 24**. (A) The discharge curve of a CR1620 Lithium coin cell battery rated at 68 mAh nominal capacity under continuous 1 mA discharge and pulsed discharge.[20] Significant capacity fade is observed for the battery with pulsed discharge. (B) Zoomed in view of the current and voltage change of the CR1620 battery.

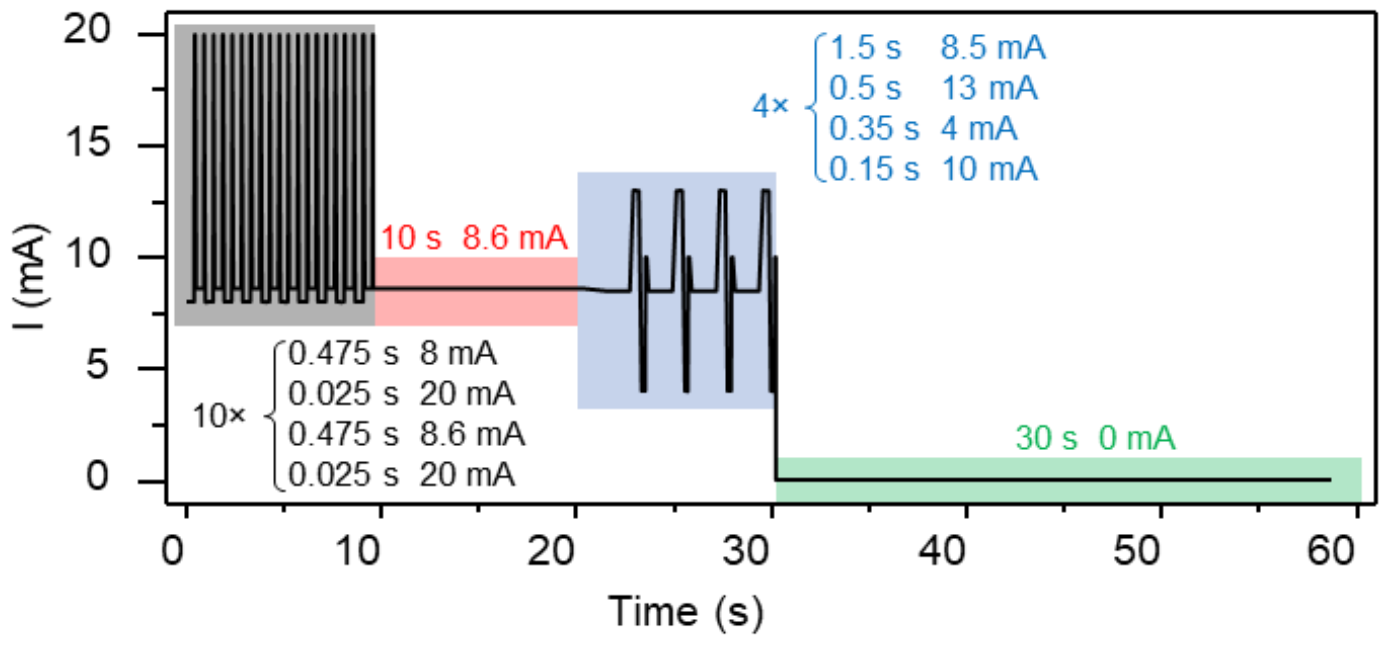

**Supplementary Figure 25**. The detailed breakdown of the pulsed discharge profile.

**Supplementary Videos**

Video S1: Repeated bending of a 1 × 5 $cm^2$ printed battery using a linear motor. (5x speed)

Video S2: The powering of the flexible E-ink display system using two printed battery connected in series.

Video S3: The powering of an LED bulb using a 1 × 5 $cm^2$ battery during various deformations.

Video S4: Low resolution micro-CT scan of the bent 1 × 5 $cm^2$ battery.

Video S5: High resolution micro-CT scan of a zoomed-in segment of the bent 1 × 5 $cm^2$ battery.

## Supplementary References